\documentclass{aa}  

\usepackage{graphicx,url}
\usepackage[normalem]{ulem}
\usepackage{mathrsfs,amssymb,amsmath}
\usepackage[varg]{txfonts}
\usepackage{booktabs}
\usepackage{multirow}
\usepackage[breaklinks, colorlinks, citecolor=blue]{hyperref}
\def \HI{H\,{\sc i}}
\def \Planck{{\it Planck}}
\def \Q{{\it Q}}
\def \U{{\it U}}

\begin{document} 

\definecolor{andrea}{rgb}{0.1, 0.0, 0.7}
\definecolor{raphael}{rgb}{0.7, 0.1, 0.5}
\newcommand{\PolFraction}{$p_{353}$}
\newcommand{\DustAngle}{$\psi_{353}$}
\newcommand{\HIAngle}{$\psi_{\rm HI}$}
\newcommand{\AngleDifference}{$\psi_{\rm 353/HI}$}
\newcommand{\secref}[1]{Sect.~\ref{#1}}
\newcommand{\figref}[1]{Fig.~\ref{#1}}
\renewcommand{\eqref}[1]{Eq.~\ref{#1}}
\newcommand{\appref}[1]{App.~\ref{#1}}

   \title{The multiphase interstellar medium as a common origin for magnetic misalignment and $TB$ parity violation}
   
\titlerunning{The multiphase ISM as a common origin for magnetic misalignment and $TB$ parity violation}
\authorrunning{Bracco et al.}

\author{Andrea Bracco\inst{1,2,3}
\and Ari J. Cukierman\inst{4}
\and Raphael Skalidis\inst{5, \thanks{Hubble Fellow}}
\and François Boulanger\inst{3}
        }

\institute{
{LUX, Observatoire de Paris, Université PSL, Sorbonne Université, CNRS, 75014 Paris, France\\
\email{andrea.bracco@obspm.fr}}
\and
INAF – Osservatorio Astrofisico di Arcetri, Largo E. Fermi 5, 50125 Firenze, Italy
\and
Laboratoire de Physique de l'Ecole Normale Sup\'erieure, ENS, Universit\'e PSL, CNRS, Sorbonne Universit\'e, Universit\'e de Paris, F-75005 Paris, France
\and
Department of Physics, California Institute of Technology, Pasadena, CA 91125, USA
\and
TAPIR, California Institute of Technology, MC 350-17, Pasadena, CA 91125, USA
 \\
         }

   \date{Accepted: 21/05/2026}

 
  \abstract
   {We present an original data analysis and a physical model that provide new insights into the origin of and relationship between two observables of the dusty polarized Galaxy at intermediate and high latitudes:~(i)~the misalignment between \HI~filamentary structures and magnetic fields and (ii)~the positive $TB$~correlation measured in \Planck\ data suggesting parity violation in the interstellar medium~(ISM). 
   We confirm an observational link between the two effects and find that both are predominantly produced at large angular scales~($\geq 10^\circ$, multipoles~$\ell \leq 20$), with a significantly stronger signal in the northern hemisphere. 
   
   We propose a model in which filaments and magnetic fields appear misaligned in projection because they are sourced by cold and warm gas phases distributed in different proportions in the solar neighborhood, from the wall of the Local Bubble to larger distances. These projection effects at large angular scales can produce coherent signatures that propagate to smaller scales in power spectra without invoking local, small-scale filament misalignment. Within this frame, \HI~filaments remain statistically aligned with the magnetic field in~3D, although with a projected scatter of tens of degrees that requires further investigation. 

   The multiphase, geometrical model presented in this work is supported by \Planck\ polarization data at 30~GHz, where synchrotron radiation dominates, and at 217 and 353~GHz, where dust emission dominates. Our analysis also incorporates starlight polarization measurements. The model introduced here suggests a new interpretation of two unexplained observables and emphasizes the role of the large-scale magnetized ISM in shaping polarized Galactic emission, which has important implications for both Galactic astrophysics and cosmological foreground characterization.
}

   \maketitle
%

\section{Introduction}\label{sec:intro}

The filamentary interstellar medium~(ISM) has recently emerged as a key research focus in the study of dusty and gaseous density structures that feed stellar nurseries in the Galaxy~\citep[e.g.,][]{Andre2010, Hacar2023, Pineda2023}. It also plays an important role in the characterization of Galactic foregrounds affecting cosmological signals in the context of cosmic microwave background~(CMB) polarization experiments~\citep[e.g.,][]{Clark2015, PlanckXXXVIII2016, PlanckXI2020, Clark2021, Cukierman2023, Halal2024, HerviasCampos2025}.

Polarized dust emission, which is produced by paramagnetic grains aligning with magnetic fields~\citep[e.g.,][references therein]{Hoang2018}, is a dominant foreground component for cosmological studies of the submillimeter sky. Dust emission obstructs accurate measurements of CMB polarization statistics, which are usually expressed in terms of angular power spectra~\citep{Bicep2_2015, PlanckXXX2016}. 
These spectra are the standard tools for quantifying the cosmological, scale-dependent amplitudes of temperature (total intensity) anisotropies~($T$) as well as parity-even~($E$) and parity-odd~($B$) modes of polarized anisotropies generated at the last-scattering surface in the early Universe~\citep[e.g.,][]{Zaldarriaga2001}.

While parity symmetry holds in the standard cosmological model, leading to vanishing $TB$ and $EB$~correlations~\citep{Zaldarriaga1997}, nonstandard theories suggest the possibility of parity violation, either during inflation~\citep[e.g.,][]{Lue1999} or through the interaction of CMB photons with parity-violating pseudo-scalar fields~(e.g., axions) during cosmic expansion~\citep[e.g.,][]{Komatsu2022}. These violations could leave observable signatures, such as cosmic birefringence, producing nonzero $TB$ and $EB$~cross-spectra.
These cosmological signatures remain undetected, with current studies providing only upper limits~\citep[e.g.,][]{PlanckXLIX2016, Eskilt2022}. In addition to the technological challenge of achieving the sensitivity and systematic control required for precision polarization measurements \citep[e.g.,][]{Ritacco2024}, potential sources of $TB$~correlation may also arise from Galactic polarized foregrounds.
A nonzero $TB$~signal has been measured in the Milky Way by the \Planck\ satellite at 353~GHz~\citep{PlanckXI2020}. This signal is primarily detected at large angular scales, corresponding to multipoles of~$\ell \leq 500$, and it has also been confirmed at 23~GHz in data from the Wilkinson Microwave Anisotropy Probe~\citep[WMAP,][]{Weiland2020}. Despite these detections, a clear physical explanation for the Galactic $TB$~correlation remains elusive.

Two main hypotheses have been proposed. 1)~The nonzero $TB$~correlation may arise from a misalignment of filamentary density structures and their local magnetic field~\citep[e.g.,][]{Huffenberger2020, Clark2021, HerviasCampos2022, Cukierman2023, HerviasCampos2025}. 2)~As proposed by \citet{Bracco2019a} and further explored by \citet{Weiland2020}, the $TB$~correlation may be imprinted by the large-scale structure of the magnetized ISM in the solar neighborhood without relying on the local small-scale magnetic misalignment.

The misalignment hypothesis is supported by observational evidence linking, in projection, filamentary density structures to the morphology of the magnetic field in the ISM. Based on dust polarization data, interstellar magnetic fields have been found to be statistically aligned with density structures in the diffuse ISM~(with column densities~$\leq 10^{22}$ cm$^{-2}$), becoming progressively perpendicular in denser molecular-cloud regions~\citep{PlanckXXXII2016, PlanckXXXV2016}. The relative orientation between filamentary structures and magnetic fields naturally induces cross-correlations among $T$, $E$, and $B$~modes. As predicted by \citet{Zaldarriaga2001} and measured in several analyses of \Planck\ data, the observed variation in relative orientation produces a positive $TE$~correlation in the diffuse ISM~\citep{PlanckXXXVIII2016} and a vanishing~$TE$ in molecular clouds~\citep{Bracco2019b}. In this framework, the $TB$~correlation could result from a statistical oblique misalignment between filamentary density structures and the local magnetic-field orientation. A misalignment angle on the order of a few degrees has been measured by comparing the morphology of the diffuse density structure, traced by atomic hydrogen~(\HI), with the magnetic field, traced by \Planck\ polarization~\citep{Clark2015, Clark2021, Cukierman2023, Halal2024}. These results support the hypothesis that magnetically misaligned filamentary structures could contribute to the nonzero $TB$~correlation measured by \Planck. However, the mechanisms inducing this misalignment remain unknown. The coherent oblique misalignment on the sky is surprising given that turbulence dynamics relax to configurations where density structures align parallel or perpendicular to the magnetic field~\citep{Soler2017}.

The large-scale-structure hypothesis addresses the issue of coherence by invoking the specific viewpoint of the solar neighborhood within the larger Galaxy. We note that in \citet{Bracco2019a}, the $TB$~correlation was modeled only for multipoles $\ell < 25$, whereas \Planck\ results show a nonzero $TB$~signal even at higher multipoles. This apparent contradiction could be resolved by recognizing that dust polarization observations do not continuously sample the large-scale magnetic field structure. Due to the sparse ISM density distribution, even a low-multipole~(large-scale) feature in the magnetic-field structure could be modulated to higher multipoles~(smaller scales). However, this large-scale picture has yet to provide a physical explanation for the origin of the misalignment.

In this work, we present a novel analysis of both the misalignment angle and the $TB$~correlation, and we introduce a physical scenario that builds on the two aforementioned hypotheses. Our interpretation is based on purely geometrical arguments in the context of the multiphase, magnetized ISM in the solar neighborhood. In the following, we introduce our assumptions and briefly describe the physical scenario. 
\begin{itemize}
\item The polarized sky at high Galactic latitudes is determined by the combination of the magnetic field of the Local Bubble~(LB), a hundreds-of-parsec cavity around the Sun carved by several supernovae~\citep{Pelgrims2020, Zucker2022}, and the Galactic mean field at larger scales.

\item We assumed the gas at high latitudes is a two-phase medium composed of a mass-weighted cold (CNM) and volume-filling warm (WNM) neutral media \citep{Wolfire2003}. Dust polarization traces both components, while the {\HI}-filamentary structures, obtained through local spatial filtering of spectroscopic data, only trace the CNM \citep[see also,][]{Clark2019}. 

\item The CNM filaments are considered compressed structures on the surface of the LB \citep[e.g.,][]{Inoue2016},  contributing to the majority of the $T$ morphology of dust intensity, and are statistically aligned with the LB magnetic field. 

\item The observed misalignment angle, measured as a large-scale effect~($\ell < 20$), is the result of projection effects along the line of sight between the LB field, traced by the CNM, and its superposition with the mean field in the spatially larger WNM. A different morphology of the LB field with respect to the mean field \citep[e.g.,][]{Alves2018} can imprint large-scale polarization structures, sourcing both the sign of the misalignment and the $TB$~correlation. 
\end{itemize}

We support this scenario using multiwavelength, high-Galactic-latitude observations in polarization including \HI~data; \Planck\ data at 30, 217, and 353~GHz; and starlight polarization measurements. The paper is organized as follows. In \secref{sec:data}, we describe the multiple datasets used in the analysis. In \secref{sec:methods}, we detail the measurement of the misalignment angle. In \secref{sec:res}, we present the main observational results of the paper, namely the dependence of the misalignment angle on sky fraction, angular scale, and polarization fraction. In \secref{sec:discussion}, we discuss the results and introduce the geometrical interpretation of the misalignment angle and the $TB$~correlation. We conclude in \secref{sec:sum} and include five appendices.     

\section{Data}\label{sec:data}
In this section, we describe the various datasets used in the analysis. Both \HI\ and \Planck\ data are presented.

\subsection{\Planck\ data}\label{ssec:planck}

We employed \Planck\ polarization data at three different frequencies:~30, 217, and 353~GHz. The higher frequencies are dominated by thermal emission from dust, while the 30-GHz data trace non-thermal synchrotron radiation. In the case of dust frequencies, we considered two sets of Stokes~\Q\ and \U\ maps~(hereafter, \Q$_{k,\rm d}$ and~\U$_{k,\rm d}$, where the subscript "$k$" represents the frequency index and the subscript "d" indicates dust) produced by different processing methods, which both improved systematic effects in polarization of {\it Planck} data compared to the legacy products of the public release~(PR)~3~\citep{planckI2020}. 

First, we used the {\tt{SRoll2}} maps~\citep{Delouis2019}. The dominant systematic effect for the polarized signal at 353~GHz in the PR3 maps is related to the measurement of the time transfer function of the detectors; at lower frequencies, it is the non-linearity of the analog-to-digital converters. Both systematics have been improved in a consistent way with all other known effects for the {\tt{SRoll2}} dataset. Second, we used the PR4 maps produced with the {\tt NPIPE} processing pipeline~\citep{PlanckLVII2020}, whose main improvements over PR3 are lower noise and systematics as well as greater internal consistency among the frequency channels.

At 353~GHz, we used the Stokes~$I$ map produced with the Generalized Needlet Internal Linear Combination~(GNILC) method~(hereafter,~$I_{\rm d}$), which is corrected for fluctuations of the CMB and the cosmic infrared background~(CIB). However, the value of the CIB monopole that must be subtracted is 0.13~MJy sr$^{-1}$ as reported in \citet{PlanckXII2020}; this is crucial to correctly estimate the dust polarization fraction, especially at high latitudes, where the Galactic emission becomes dimmer. 
We denote polarization fraction by 
\begin{equation}\label{eq:p}
  p_{\rm d} = P_{\rm d}/I_{\rm d} = \sqrt{Q_{353,\rm d}^2+U_{353,\rm d}^2}/I_{\rm d},  
\end{equation}
where $P_d$~is referred to as the polarized intensity. In order to convert units from~K$_{\rm CMB}$ to~MJy sr$^{-1}$ at 353~GHz, we used the conversion factor~287.5~\citep{PlanckXII2020}. 
\begin{figure}[!t]
\begin{center}
\resizebox{1\hsize}{!}{\includegraphics{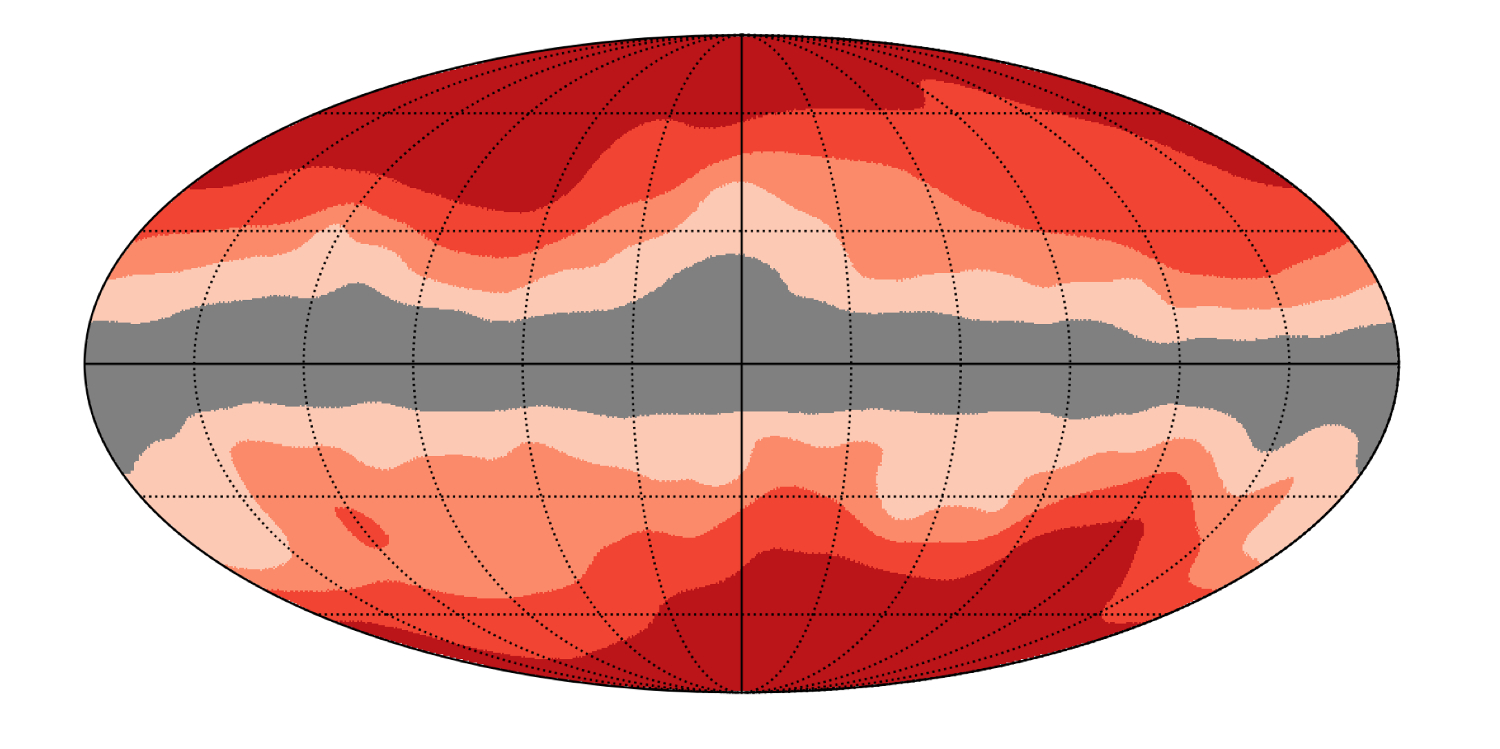}}
\caption{All-sky masks with sky fractions of~20\%, 40\%, 60\%, and~80\% with progressively brighter colors. The gray region along the Galactic plane is not considered in this work. A galactic coordinate grid with steps of~30$^{\circ}$ in longitude~$l$ and latitude~$b$ is overlaid with its origin at the Galactic center.} \label{fig:masks}
\end{center}
\end{figure}
\begin{figure*}[!h]
\begin{center}
\resizebox{0.9\hsize}{!}{\includegraphics{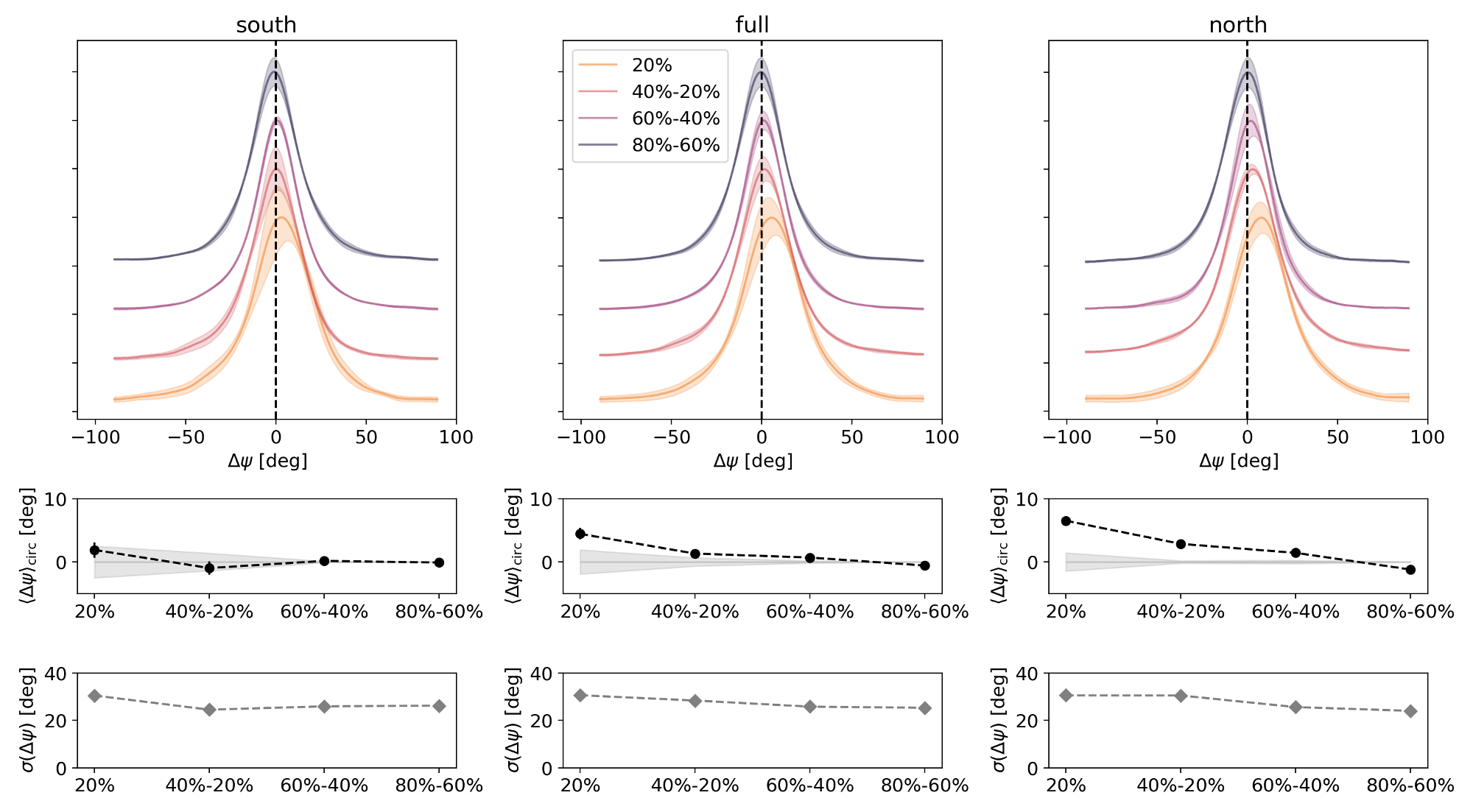}}
\caption{Normalized probability distribution functions of~$\Delta \psi$ between \Planck\ data at 353~GHz and the \HI~templates. The value of $\Delta \psi$ is computed for different sky masks at angular and pixel resolutions of~80$\arcmin$ and $N_{\mathrm{side}} =128$, respectively. The NPDFs are shown for both Galactic hemispheres combined~(central panels) and for the southern~(left panels) and northern~(right panels) hemispheres individually. The corresponding circular-mean values~(in black circles) and their standard deviations~(in gray diamonds) are shown in the insets at the bottom as a function of the sky mask. The gray shaded area in the middle panels represents the misalignment angle caused by systematic differences between {\tt{SRoll2}} and PR4 data.} 
\label{fig:Dpsi_masks}
\end{center}
\end{figure*}
\begin{figure}[!h]
\begin{center}
\resizebox{0.8\hsize}{!}{\includegraphics{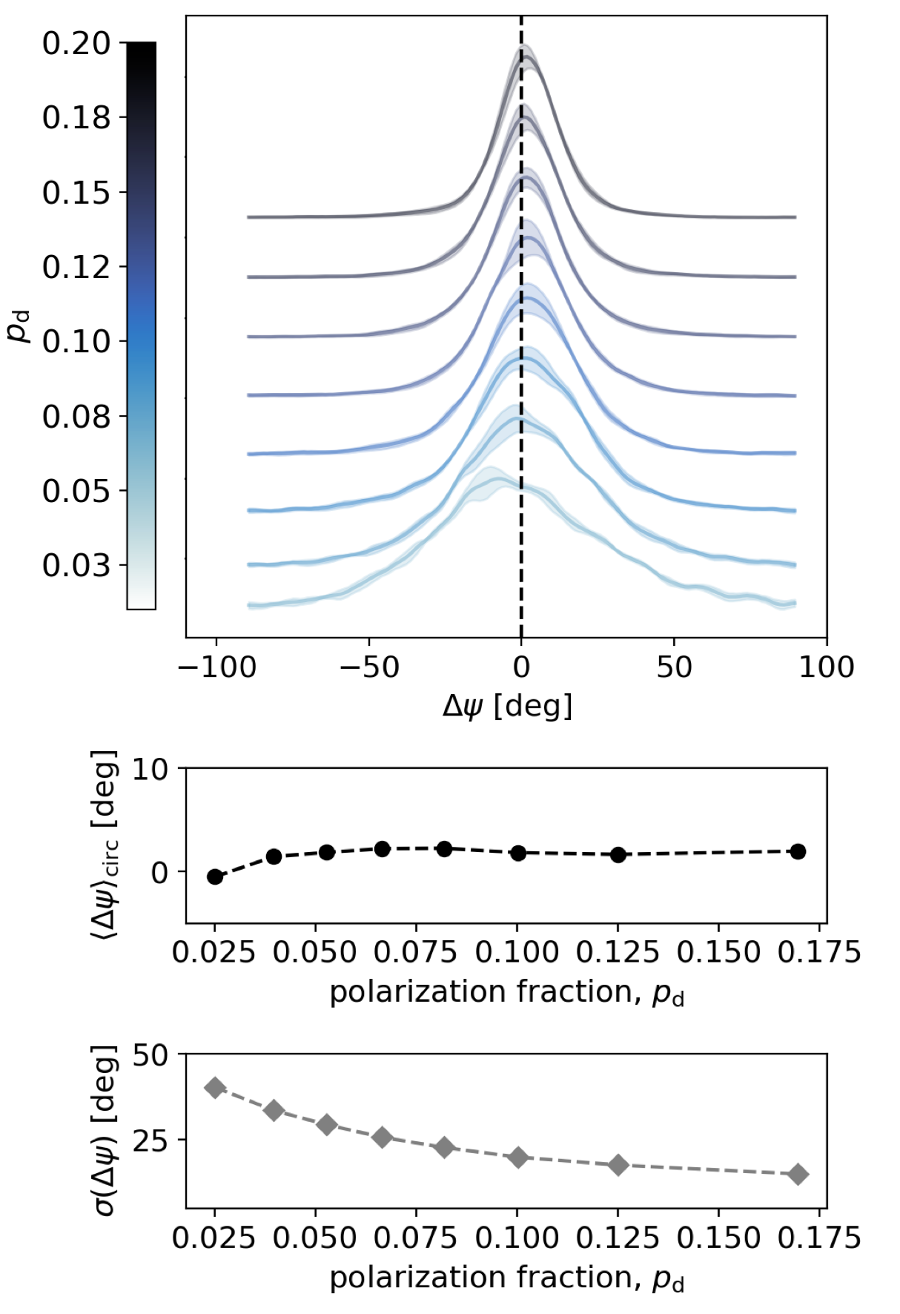}}
\caption{The NPDFs of~$\Delta \psi$ as a function of~$p_{\rm d}$ over the 80\%~mask. The NPDFs were computed in equally sampled bins of~$p_{\rm d}$ containing 15633~elements and are represented with bright to dark colors from low to large values of~$p_{\rm d}$, respectively. The central value of each~$p_{\rm d}$ bin is plotted on the $x$~axis in the central and bottom panels. The errors on~$\langle\Delta \psi\rangle_{\rm circ}$ are smaller than the symbols.} 
\label{fig:Dpsi_pd}
\end{center}
\end{figure}
In the case of the Stokes parameters at 30~GHz (hereafter, \Q$_{\rm s}$ and~\U$_{\rm s}$, where the subscript "s" indicates synchrotron radiation), we considered two distinct datasets from PR3 and PR4, respectively. 
In the following, all maps are in {\tt HEALPix}\footnote{\url{http://HEALPix.sf.net}} format~\citep{Gorski2005}. 
The data of reference are~\Q$_{353,\rm d}$ and~\U$_{353,\rm d}$ because dust polarized emission is maximum there. In order to neglect noise bias in polarization over the full sky at 353~GHz, the best angular resolution of all maps is a full-width half-maximum~(FWHM) of~80$\arcmin$  \citep{PlanckXII2020}. For this FWHM, the corresponding pixel resolution is determined by the {\tt HEALPix} parameter $N_{\mathrm{side}} = 128$~($\sim 30\arcmin$~pixel width). All maps are smoothed and projected on the same sky grid using the {\tt healpy} Python package~\citep{Zonca2019}. 

\subsection{\HI\ data}\label{ssec:hi}

We used two polarization templates derived from maps of~{\HI} emission\footnote{\url{https://dataverse.harvard.edu/dataset.xhtml?persistentId=doi:10.7910/DVN/74MEMX}}. Each technique identifies the orientations of filamentary \HI~structures and infers a plane-of-sky magnetic-field orientation, which implies a perpendicular dust polarization angle. The two techniques are described in sect.~3 of \cite{Halal2024}, and the \HI~maps are drawn from the spectroscopic data cubes of \cite{HI4PI2016} with velocities between~-13 km s$^{-1}$ and~16~km~s$^{-1}$. One template is formed from the Hessian matrix of the \HI~brightness temperature; the polarization angle is based on locations of negative curvature, and the associated polarization intensity is based on the Hessian eigenvalues. The second template is based on the spherical rolling Hough transform~(SRHT), which weights linear structures according to their local orientations; this allows for a superposition of orientations, and the polarization intensity is based on the \HI~brightness temperature. When comparing the two \HI~maps~(hereafter, ``\HI~templates'') with the \Planck\ polarization data, we smoothed them to the same FWHM and pixel resolution.

\subsection{Starlight polarization catalog}\label{ssec:stars}

In \secref{sssec:stars}, we study the dependence of polarization-angle differences on heliocentric distance. For this analysis, we use the most recently compiled starlight polarization catalog with distance estimates from {\it Gaia} \citep{Panopoulou2025}. Starlight polarization provides complementary information to the dust thermal emission, as it is the result of extinction on dust grains along the line of sight and allows one to trace the magnetic field orientation averaged from the observer to the stars \citep[][]{Hiltner1949, Davis1951, Hildebrand1988}. We stress that, while dust polarization in emission traces the orthogonal orientation to the magnetic-field on the plane of the sky, starlight polarization directly traces the magnetic-field orientation. 

We considered stars with a polarization angle uncertainty smaller than~5$^{\circ}$~(or a signal-to-noise in degree of polarization on the order of~5). We projected the star catalog onto {\tt HEALPix} grids with $N_{\rm side} = 128$, averaging over any multiple starlight measurements within the same {\tt HEALPix} pixel~(see \secref{sssec:stars}). We verified that our results are robust to changes in the pixelation.

\section{Methods}\label{sec:methods}

We computed the relative orientation between two sets of polarization angles~$i$ and~$j$, namely, $\psi_i = 0.5\times\,{\rm \tt atan2}(U_{i}, Q_{i})$ and $\psi_{j} = 0.5\times\,{\rm \tt atan2}(U_{j}, Q_{j})$, as follows, 
\begin{equation}\label{eq:Dpsi}
    \Delta \psi_{ij} = \frac{1}{2}{\rm \tt atan2} (\mathcal{A}_{ij}, \mathcal{B}_{ij}), 
\end{equation}
where $\mathcal{A}_{ij} = (\sin{2\psi_{i}}\cos{2\psi_{j}} - \cos{2\psi_{i}}\sin{2\psi_{j}})$ and $\mathcal{B}_{ij} = (\cos{2\psi_{i}}\cos{2\psi_{j}} + \sin{2\psi_{i}}\sin{2\psi_{j}})$ \citep[see also,][]{PlanckXXXII2016, Clark2015}.
We note that, in this work, the sign of the polarization angles is positive for consistency with the IAU convention and the notation used in \citet{Cukierman2023}; following the {\tt HEALPix} convention, however, a minus sign would appear in \eqref{eq:Dpsi}~\citep[e.g.,][]{PlanckXLIV2016}.

In \secref{ssec:powerspectra_data}, as part of the link between $TB$ correlation and misalignment angle, we form smoothed versions of the polarization maps in order to estimate the polarization angles associated with large-scale features. We use the large-scale polarization angles to de-rotate the original, unsmoothed maps. This has the effect of removing the prevailing orientations of large-scale polarization features. The smoothed map is determined by a given FWHM represented by the multipole~$\ell_{\rm ref}$, i.e., FWHM = $180^\circ/\ell_{\rm ref}$. The rotated Stokes parameters~(labeled with the superscript "R") were obtained using the following rotational transform: 
\begin{equation}\label{eq:rot}  
    \begin{pmatrix} Q_{x}^{\rm R} \\ U_{x}^{\rm R} \end{pmatrix} = \left(\begin{matrix} \cos{2\psi_{x,\ell_{\rm ref}}} & \sin{2\psi_{x,\ell_{\rm ref}}} \\ -\sin{2\psi_{x,\ell_{\rm ref}}} & \cos{2\psi_{x,\ell_{\rm ref}}}\end{matrix} \right) \begin{pmatrix} Q_{x} \\ U_{x} \end{pmatrix},
\end{equation}
where $x$~denotes an index that can be either~$i$ or~$j$. As an example, in \figref{fig:stokes} the case for $\ell_{\rm ref} = 20$ is shown applied to~\Q$_{353,\rm d}$ and~\U$_{353,\rm d}$. 

In \secref{ssec:fsky}, we study how the histograms of~$\Delta \psi$~, normalized probability distribution functions (NPDFs), vary across the sky. In particular, we made use of four distinct masks delivered by the \Planck\ Collaboration\footnote{\url{https://pla.esac.esa.int/}}, which cover progressively larger portions of the sky, ranging from~20\% to~80\% with steps of~20\%. These masks are produced by masking out incrementally brighter thermal dust emission. Figure~\ref{fig:masks} displays these masks in different colors on a galactic coordinate grid. The masks approximately correspond to selections of sky areas by Galactic latitude, such that the 20\%~mask mostly includes regions at $|b| > 60^\circ$. Averaging the NPDFs obtained by using Eq.~\ref{eq:Dpsi} among all distinct data versions (i.e., {\tt{SRoll2}}, PR4, SRHT, Hessian), in the following we show their mean distribution, $\Delta \psi$, and the corresponding standard deviation. The impact of residual systematic effects in the \Planck\ data on the NPDFs is quantified by calculating the angle difference between the {\tt{SRoll2}} and PR4 datasets and is shown as a gray shaded area in the figures.

In \secref{ssec:scale}, focusing on the 20\% mask, we explore how the NPDFs of~$\Delta \psi$ depend on angular scale in the two Galactic hemispheres. We computed~$\Delta \psi$ after smoothing both the \Planck\ data and the \HI~templates to progressively lower angular resolutions, parameterized by the multipole $\ell_{\rm FWHM} = 180/{\rm FWHM\ [deg]}$. To avoid leakage from bright emission at low Galactic latitudes, we applied the 40\%~mask to the Stokes parameters before smoothing. We further applied a positive (negative) Galactic latitude criterion to the 40\%~mask to isolate the northern (southern) hemisphere. The masks were also smoothed with a Gaussian beam to minimize artifacts introduced by the smoothing process.

Throughout the analysis, we maintained a fixed pixelization, independent of the smoothing kernel. While this resulted in oversampling of the beam at low angular resolution, it allowed us to retain the same number of sky pixels in the $\Delta \psi$~NPDFs. We verified that the results remained robust when re-pixelizing the maps to lower values of~$N_{\mathrm{side}}$, ensuring they remained within the Nyquist-sampling limit. 
In \secref{ssec:powerspectra_data}, in order to explore the impact of large-scale features on the dust polarization power spectra at small scales, we use the \Planck\ and \HI-template maps at FWHM=15$\arcmin$ with $N_{\mathrm{side}} = 512$.

\section{Results}\label{sec:res}

In this section, we report the observational results on the angle difference~$\Delta \psi$ between \Planck\ polarization data at 353~GHz and the \HI~templates. We present the dependence of~$\Delta \psi$ on sky fraction, on~$p_{\rm d}$ and on angular scale.

\subsection{Misalignment as a function of sky fraction and $p_{\rm d}$}\label{ssec:fsky}

By applying \eqref{eq:Dpsi}, we computed~$\Delta \psi$ and corresponding NPDFs for combined and separate Galactic hemispheres using data at FWHM=80$\arcmin
$. In \figref{fig:Dpsi_masks}, we show the NPDFs of~$\Delta \psi$, their circular means\footnote{The circular, or angular, mean is a directional statistic that differs from a standard mean as it applies to cyclical quantities.}~$\langle\Delta \psi\rangle_{\rm circ}$ and their spread~$\sigma(\Delta \psi)$ as a function of sky fraction. The central panels include gray shadows representing~$\Delta \psi_{\rm circ}$ values computed using the difference between {\tt{SRoll2}} and PR4 polarization data, a measure of residual systematic effects in the \Planck\ data. In the top panels, errors on the NPDFs reflect variance across different versions of the same dataset used for each tracer, namely, dust polarization and \HI~templates.

Firstly, we observe that all NPDFs exhibit a large value of~$\sigma(\Delta \psi)$, which is approximately~$30^\circ$ in both Galactic hemispheres. The only significant variation of~$\sigma(\Delta \psi)$, of about a factor of three, is found with respect to~$p_{\rm d}$. In \figref{fig:Dpsi_pd}, we show the NPDFs of~$\Delta \psi$ as a function of nine equally sampled bins of~$p_{\rm d}$. Each bin contains 15633~sky pixels. We verified that results are robust to changes in the number of bins. The central value of each $p_{\rm d}$~bin is plotted on the $x$~axis in the central and bottom panels of the figure. The corresponding NPDFs are shown in the top panel with brighter~(darker) colors for lower~(larger) values of~$p_{\rm d}$. As shown by the bottom panel, $\sigma(\Delta \psi)$~varies from~$40^\circ$ to~$15^\circ$, spanning from low to large values of~$p_{\rm d}$ in the 80\%~mask. This result shows that $\sigma(\Delta \psi)$~is the lowest when $p_{\rm d}$ is the largest. As $p_{\rm d}$~is maximum when the magnetic field is perpendicular to the line of sight~\citep[e.g.,][]{PlanckXLIV2016}, the trend of $\sigma(\Delta \psi)$ that we observed is most likely determined by projection effects, i.e., \HI\ templates more closely align with dust polarization when the magnetic field is perpendicular to the line of sight~\citep[see also,][]{Clark2019b}. We notice, however, that at FWHM=80$\arcmin$ even for the largest $p_{\rm d}$ values, $\sigma(\Delta \psi)$ is yet larger than $10^{\circ}$, highlighting a significant dispersion around the alignment.   

Secondly, we observe that although the NPDFs generally peak at $\langle\Delta \psi\rangle_{\rm circ} \approx 0^\circ$, this is not the case for the 20\%~mask at high Galactic latitudes, where a misalignment of $4.5^\circ \pm 0.9^\circ\footnote{This is the standard error on the mean.}$ occurs on average across the two hemispheres, in agreement with previous findings~\citep[e.g.,][]{Clark2021, Cukierman2023}. This misalignment is more prominent in the northern hemisphere, reaching a value of $\langle\Delta \psi\rangle_{\rm circ} = 6.5^\circ \pm 0.6^\circ$, significantly exceeding the contributions from systematic effects. In the southern hemisphere the misalignment is always consistent with systematic effects. In the northern Galactic hemisphere, we notice that the misalignment remains significant in the 40\%~mask at lower Galactic latitudes. In the central panel of \figref{fig:Dpsi_pd}, we also observe that the misalignment is independent of~$p_{\rm d}$ for values larger than~0.03. The level of misalignment in the 80\%~mask is about~$2^\circ$ and drops to~$0^\circ$ for $p_{\rm d} < 0.03$, where data noise is likely contributing to the NPDFs. We stress that the value of~$\sim$2$^\circ$ is not comparable with what is shown in \figref{fig:Dpsi_masks}, where we computed the misalignment angle averaged in the 20\%~of the sky between the 80\% and the 60\%~masks.

\begin{figure*}[!h]
\begin{center}
\resizebox{0.8\hsize}{!}{\includegraphics{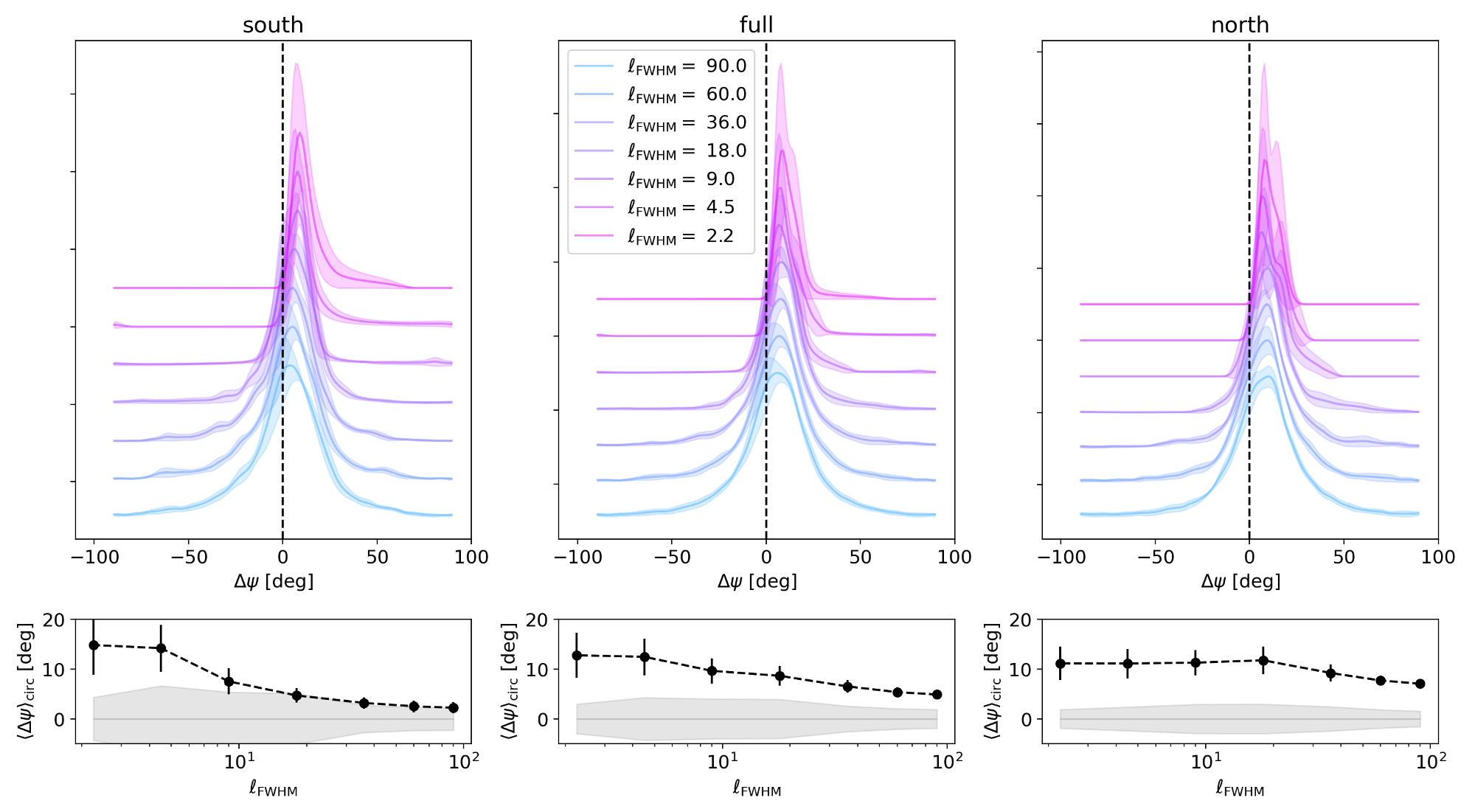}}
\caption{The NPDFs of $\Delta \psi$, considering the 20\%~mask and changing the angular resolution of the maps to~$\ell_{\rm FWHM}$. Left and right panels show the southern and northern hemispheres, respectively. The central panel shows them together. More small-scale information is included as $\ell_{\rm FWHM}$~increases.} 
\label{fig:Dpsi_scale}
\end{center}
\end{figure*}

\begin{figure*}[!h]
\begin{center}
\resizebox{0.8\hsize}{!}{\includegraphics{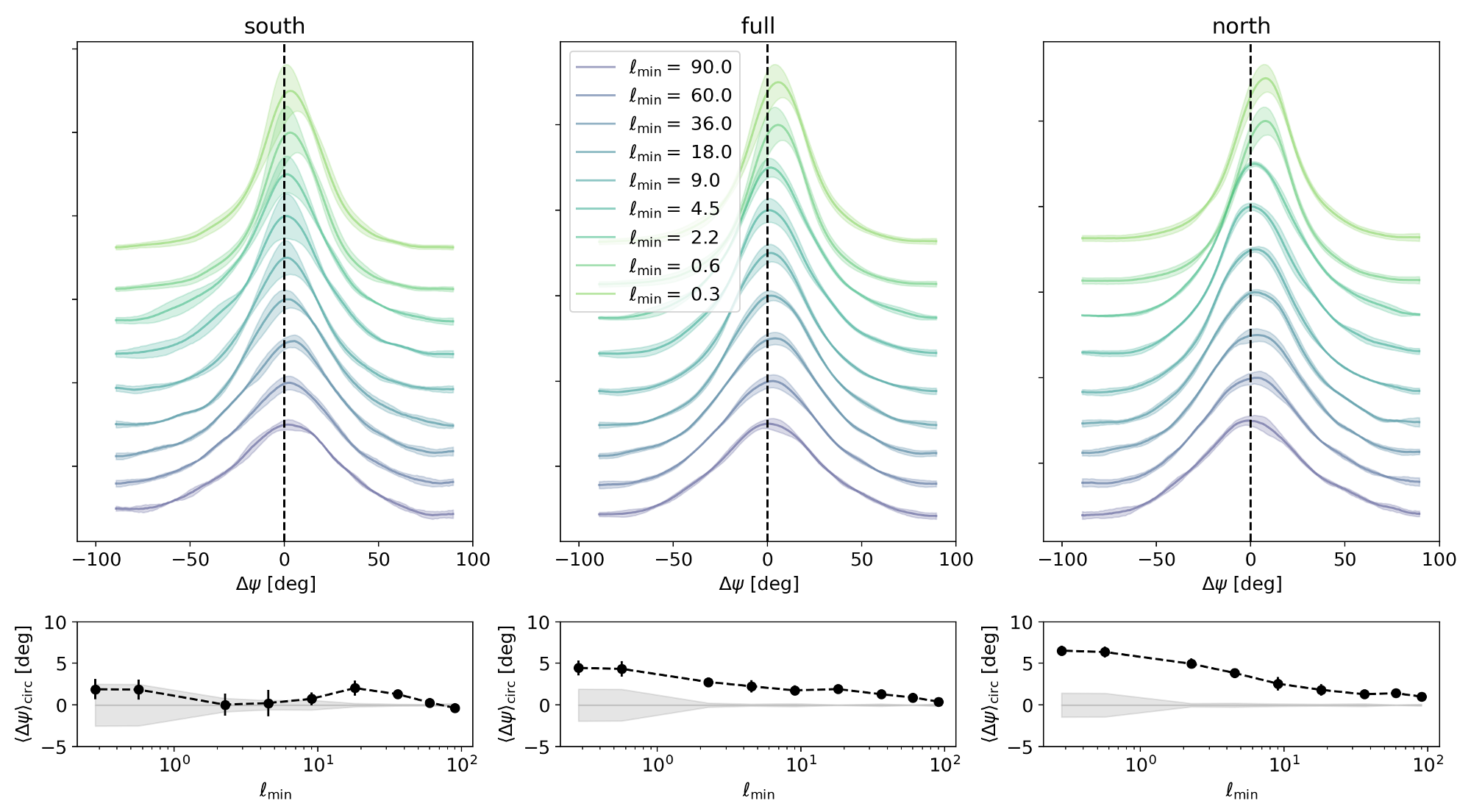}}
\caption{Same as for \figref{fig:Dpsi_scale} but filtering out the large angular scales up to~$\ell_{\rm min}$. Large-scale information is removed at $\ell < \ell_{\rm min}$ as $\ell_{\rm min}$~increases.} 
\label{fig:Dpsi_scale_prog}
\end{center}
\end{figure*}

\subsection{Misalignment as a function of scale}\label{ssec:scale}

Using the 20\%~mask, we explored the dependence of~$\Delta \psi$ on angular scale in two separate but complementary ways. 
First, we computed~$\Delta \psi$ after smoothing the \Planck\ data and the \HI~templates to progressively lower values of~$\ell_{\rm FWHM}$.  In \figref{fig:Dpsi_scale}, we show the NPDFs of~$\Delta \psi$ as a function of~$\ell_{\rm FWHM}$. We observe an increase in misalignment at large angular scales, while $\sigma(\Delta \psi)$~remains uniform, suggesting that data noise is not the dominant contribution to~$\sigma(\Delta \psi)$. These effects occur in both Galactic hemispheres with the same misalignment sign. The largest misalignment, approximately $12^\circ \pm 3^\circ$, occurs at $\ell_{\rm FWHM} = 4.5$. At these scales, the misalignment becomes significant compared to systematic effects, even in the southern Galactic hemisphere. We also notice that at larger multipoles~($\ell > 10$) the two hemispheres behave differently, with a rather flat and coherent misalignment in the north and a decreasing misalignment in the south, which is consistent with systematic effects for all smoothing scales.

Second, we computed~$\Delta \psi$ from high-pass-filtered Stokes parameters and \HI~templates, retaining only power up to angular scales corresponding to multipoles $\ell_{\rm min} = 180^\circ/{\rm \tt{max\_scale}}$. Using a Gaussian filter, we removed power at multipoles less than~$\ell_{\rm min}$. This analysis, shown in \figref{fig:Dpsi_scale_prog} with the respective NPDFs, confirms that the observed misalignment between \Planck\ data and \HI~templates is predominantly a large-scale phenomenon, manifesting only at the largest scales including the dipole.

\section{Discussion of large-scale misalignment}\label{sec:discussion}

In this section, we discuss the large-scale misalignment and its implications for both understanding the magnetic-field structure in the solar neighborhood and interpreting the $TB$~correlation in angular power spectra. The origin of misalignment between dust polarization and \HI~templates is investigated with two physical scenarios involving line-of-sight variations of either the dust emission or the magnetic-field structure probed by the multiphase \HI~gas. Our analysis supports the latter scenario.  

\subsection{First scenario: Variations in dust emission}\label{ssec:dust}

The knowledge of the polarized spectral energy distribution~(PSED) of interstellar dust is of primary importance both for characterizing the physical properties of dust grains in our galaxy~\citep[e.g.,][]{Guillet2018, PlanckXII2020, Reissl2020} and for estimating foreground contamination to CMB polarization~\citep[e.g.,][]{PlanckXI2020}. However, it has been shown that a thorough understanding of the dust PSED may be strongly hampered by line-of-sight changes in dust properties~\citep[e.g.,][]{Tassis2015, Skalidis2024, Mandarakas2024}. As the observed dust polarization is determined by the emission-weighted Galactic magnetic field along the line of sight~\citep[e.g.,][]{Wardle1990, Lee1985,PlanckXX2015}, any changes in dust opacity and temperature could introduce effective frequency-dependent variations of polarization fraction and angle~\citep[e.g.,][]{Tram2024}. This effect was measured in \Planck\ data \citep[e.g.,][]{Pelgrims2021, Ritacco2023}.        
Since \HI~templates are sensitive to the magnetic-field orientation but not to dust properties, the observed misalignment with \Planck\ data could be a signature of changes in dust emission~(i.e., temperature and opacity) along the line of sight. Because the misalignment is predominantly at large scales, we may witness dust emissivity changes between the solar neighborhood within a few hundred parsecs and dusty regions on larger physical scales. As this scenario would imply frequency-dependent variations of the \Planck\ polarization angle, we calculated~$\Delta \psi$ using \Planck\ data at 217~GHz. In \figref{fig:Dpsi_scale_217}, we show the NPDFs of~$\Delta \psi$ between \HI~templates and dust polarization at 217~GHz as a function of~$\ell_{\rm FWHM}$. 
Comparing the NPDFs and the corresponding values of~$\langle\Delta \psi\rangle_{\rm circ}$ with the 353-GHz results presented in \figref{fig:Dpsi_scale}, we obtain comparable misalignment at the two frequencies, which cannot be caused by known systematic effects in the \Planck\ data. Although the values of~$\langle\Delta \psi\rangle_{\rm circ}$ at 217~GHz are a few degrees lower than those at 353~GHz, they are consistent within the uncertainties. Given the level of precision, we are not able to detect significant variations of the misalignment between frequencies. This suggests that frequency decorrelation is unlikely measured by the NPDFs and that other mechanisms may be responsible for the misalignment between \HI~templates and dust polarization.    

\subsection{Second scenario: Changes in magnetic-field structure}\label{ssec:bfield}

The second scenario involves only geometric effects related to variations in the magnetic-field structure associated with different phases of the \HI~gas along the line of sight. Thermal dust emission is highly correlated with the \HI~brightness temperature at intermediate and high Galactic latitudes~\citep[e.g.,][]{Boulanger1996, Lenz2017}. The \HI~gas exists as a bi-stable medium, comprising both the CNM ($T_{\rm HI} \approx 100$~K, $N_{\rm HI} \approx 50$~cm$^{-3}$) and the WNM ($T_{\rm HI} \approx 8000$~K, $N_{\rm HI} \approx 1$~cm$^{-3}$). These components, both containing interstellar dust, carry distinct temperatures and densities, with their relative proportions varying according to the Galactic environment~\citep{Wolfire2003, Ferriere2020, Marchal2021, Marchal2024}.

Assuming that dust grains have uniform properties~(such as temperature, opacity, and alignment with the magnetic field), we hypothesize that the \HI~templates and dust polarization are influenced by distinct orientations of the magnetic field, which varies with different mixtures of CNM and WNM along the line of sight. Specifically, at high Galactic latitude, the \HI~templates, derived from local spatial filtering of the \HI~brightness temperature at velocities between~-13 and~16~km~s$^{-1}$~(\secref{ssec:hi}), are thought to predominantly trace CNM filamentary structures on the LB~surface~\citep{Clark2019}. CNM also dominates the structure of dust total intensity. In contrast, dust polarization traces the line-of-sight superposition of both CNM and WNM. CNM density structures form through thermal instability, triggered by turbulence and shock-driven large-scale compressions, within the volume-filling WNM gas. These processes also impact the magnetic-field structure in both phases~\citep[e.g.,][]{Inutsuka2015, Inoue2016}. The \HI~templates at high Galactic latitudes primarily map the magnetic field on the edges of the~LB \citep{Clark2019}. Meanwhile, dust polarization provides combined information on the LB magnetic field and the regular magnetic field in the WNM over physical scales larger than the LB~\citep[> 300 pc,][]{Oneill2024}. 
For simplicity, in this work we have assumed a two-phase, two-layer model. This approximation may underestimate the presence of additional gas components along the line of sight, such as the unstable \HI\ neutral medium (UNM), and their effects on the observed polarization. Multi-layer approaches incorporating mixtures of CNM, UNM, and WNM that fit the \Planck\ data at high Galactic latitude have been implemented by \citet{Ghosh2017} and \citet{Adak2020}. 

\begin{figure}[!h]
\begin{center}
\resizebox{0.6\hsize}{!}{\includegraphics{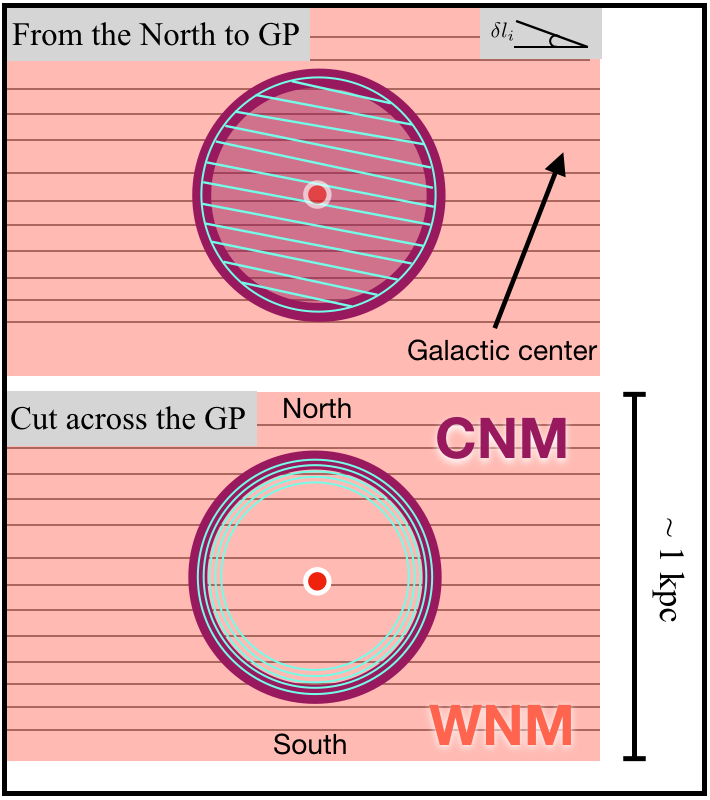}}
\caption{Sketch of the toy model showing the 
LB and the mean magnetic field in the solar neighborhood from the north pole to the Galactic plane (GP, top) and across the GP (bottom). Dark colors correspond to structures of CNM, light colors to the WNM, cyan lines to magnetic-field lines in the CNM, and purple lines to the magnetic-field lines of the WNM. The Sun position in the LB is represented by a red-white circle. An inset in the top panel shows the angle~$\delta l_i$ between the LB/CNM field and the mean/WNM magnetic field.} 
\label{fig:sketch}
\end{center}
\end{figure}

In \figref{fig:sketch}, we present a sketch of the physical scenario centered on the Sun~(red circle) both seen from the North to the Galactic plane and with a cut across it. The sketch shows the \HI~templates as being sensitive to the magnetic field in the CNM, indicated by cyan lines within the dark-purple regions, while dust polarization is sensitive to magnetic fields in both the CNM and WNM. 

Our final assumption is that the LB influences the large-scale regular magnetic field \citep{Pelgrims2025}, causing a distortion that varies between the two hemispheres. This hypothesis is supported by geometric fits to the \Planck\ data, as reported in \citet{Alves2018} and \citet{Pelgrims2020}. These studies found that the LB magnetic field consistently points towards Galactic coordinates $(l, b) = (71.0 \pm 1.3, -10.9 \pm 0.1)$ deg in the northern hemisphere and $(l, b) = (74.0 \pm 1.4, +5.8 \pm 0.7)$ deg in the southern hemisphere.

In the following section, we introduce a geometrical toy model designed to explore the origin of the large-scale misalignment. The true relative morphology of CNM and WNM magnetic fields is certainly more complex than our toy model. Our objective is not to fit the data but rather to discuss a parametric model that offers an explanation of the observed effects. 

\subsubsection{Geometrical toy model of the magnetic field}\label{ssec:modelproj}

We aim to model the misalignment angle specifically at the largest angular scales~($\ell < 20$), which are most relevant for the observed misalignment between dust polarization and the \HI~templates~(see \figref{fig:Dpsi_scale_prog}). For this purpose, we utilize the phenomenological multi-layer model introduced by \citet{PlanckXLIV2016}, which has been applied in several studies to fit the \Planck\ data at high Galactic latitudes~\citep[e.g.,][]{Vansyngel2017, PlanckL2017} and to model the rotation measure maps of the LB using the LOw Frequency ARray~\citep[LOFAR, e.g.,][]{Boulanger2024}.

In this work, we just considered the regular (ordered) magnetic field vector~$\vec{B_0}$ as we are interested in the line-of-sight geometrical variation of the magnetic field at large angular scales. We do not include any curvature term in the ordered component. Moreover, since we focus on angle differences, our model only depends on the direction of~$\hat{\vec{B}}_0$, which is assumed to be uniform and pointing toward Galactic coordinates~($l_0, b_0$), such that its coordinates are~($\cos{l_0}\cos{b_0}, \sin{l_0}\cos{b_0},\sin{b_0}$). Defining the generic
line-of-sight unit vector~$\hat{r}$ as~($\cos{l}\cos{b}, \sin{l}\cos{b},\sin{b}$), the line-of-sight~($\hat{\vec{B}}_{0,\parallel}$) and plane-of-the-sky~($\hat{\vec{B}}_{0,\perp}$) components of~$\hat{\vec{B}}_0$ can be expressed as
\begin{align}\label{eq:lospos}
\hat{\vec{B}}_{0,\parallel} &=  \hat{\vec{B}}_0 - \hat{\vec{B}}_0\cdot \hat{r}  \\\nonumber
\hat{\vec{B}}_{0,\perp} &= \hat{\vec{B}}_0 - \hat{\vec{B}}_{0,\parallel}.
\end{align}

From \eqref{eq:lospos}, we  computed the geometric parts of the Stokes parameters, $q \propto Q/I$ and $u \propto U/I$, corresponding to $\hat{\vec{B}}_{0}$ as 
\begin{align}\label{eq:qu}
q_0 &= \cos^2{\gamma_0} \cos{2\psi_0} \\\nonumber
u_0 &= - \cos^2{\gamma_0} \sin{2\psi_0}.
\end{align}
The parameter~$\gamma_0$ is the angle between the magnetic field and the plane of the sky; the parameter~$\psi_0$ is the polarization angle. These two angles are defined as
\begin{align}\label{eq:angles}
\cos^2{\gamma_0} &= 1 - (\hat{\vec{B}}_0\cdot \hat{r})^2 \\\nonumber
\psi_0 &= \pi/2 - \arccos\left({\frac{\hat{\vec{B}}_{0,\perp}\cdot \hat{n}}{|\hat{\vec{B}}_{0,\perp}|}}\right),
\end{align}
where $\hat{n}$~is the unit vector perpendicular to~$\hat{r}$ within the $\hat{r}$-$\hat{z}$~plane and $\hat{z}$~is the unit vector pointing toward the north Galactic pole in Galactic coordinates.

This toy model produces a map of~$\psi_0$ that in our scenario corresponds to the polarization angle of the regular magnetic field traced by the WNM. To introduce the impact of the LB, we imposed that, at large scales, the direction of~$\hat{\vec{B}}_{\rm LB}$ in the two hemispheres corresponds to a rotation of~$\hat{\vec{B}}_0$, such that $\hat{\vec{B}}_{\rm LB}$~points toward Galactic coordinates~($l_0 + \delta l_i, b_0 + \delta b_i$) with the index~"$i$" referring to either the southern or northern hemisphere~(see cyan lines in the LB and top-right inset in \figref{fig:sketch}). Using Eqs.~\ref{eq:lospos}, \ref{eq:qu} and~\ref{eq:angles} in the case of~$\hat{\vec{B}}_{\rm LB}$, we derived the corresponding geometrical parts of Stokes parameters~$q_{\rm LB}$ and~$u_{\rm LB}$ representing the magnetic field on the LB surface traced by the CNM.

The final step of our geometrical approach is the derivation of the total Stokes parameters, which, in terms of their geometrical parts, are modeled as 
\begin{align}\label{eq:qutot}
q_{\rm tot} &= q_{\rm LB}f_{\rm L} + q_0(1-f_{\rm L}) \\\nonumber
u_{\rm tot} &= u_{\rm LB}f_{\rm L} + u_0(1-f_{\rm L}),
\end{align}
where $f_{\rm L}$~represents the relative contribution of the LB to the total polarization signal. Finally, using \eqref{eq:Dpsi}, we computed the misalignment angle between~$\psi_{\rm tot}$, a proxy of dust polarization, and~$\psi_{\rm LB}$, a proxy of the \HI~templates, in the 20\%~mask. We notice that the strongest impact on the misalignment angle in this sky area is determined by~$\delta l_i$ compared to~$\delta b_i$. Thus, fixing $\delta b_i = 0^{\circ}$, in the models, we explored the effect on the observed misalignment angle of changing~$\delta l_i$ (the intrinsic misalignment ) and $f_{\rm L}$.

\begin{figure}[!h]
\begin{center}
\resizebox{0.8\hsize}{!}{\includegraphics{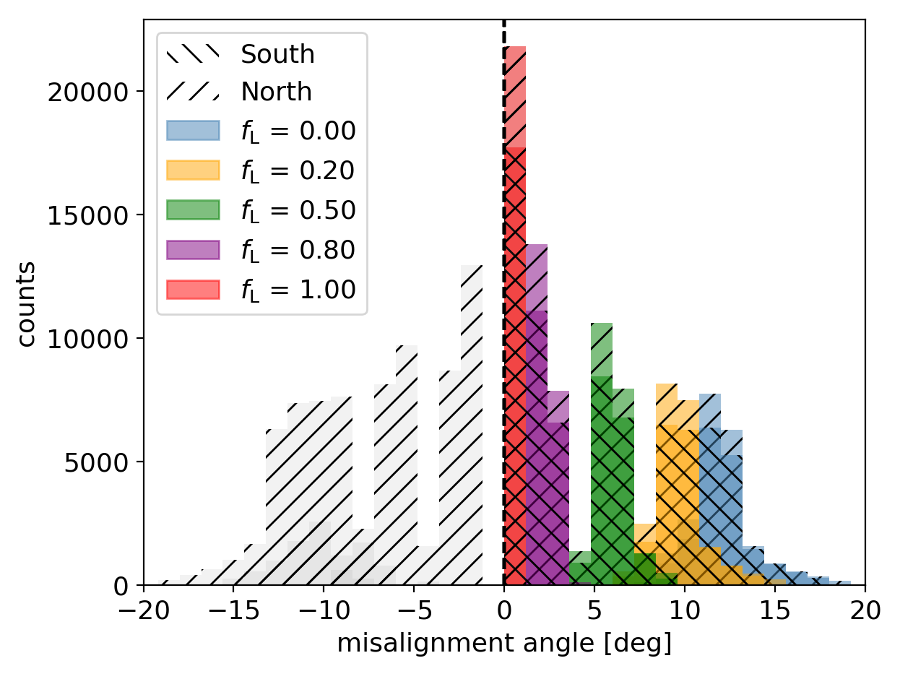}}
\caption{Modeled misalignment angle between the LB magnetic field and the total one as a function of~$f_{\rm L}$. Hatches show the misalignment in each Galactic hemisphere separately. As an example, the intrinsic misalignment angle~$\delta l_i$ is $\pm 12^\circ$ in the southern and northern hemisphere, respectively. The histograms were computed in the 20\%~mask. In light gray, the cases for both hemispheres with $\delta l_i = +12^\circ$ are shown.} 
\label{fig:toymod0}
\end{center}
\end{figure}

We fixed $(l_0, b_0)  = (72.5^\circ, -5^\circ)$ to be an intermediate direction with respect to those found by \citet{Pelgrims2020} in the southern and northern hemispheres~(see \secref{ssec:bfield}). In \figref{fig:toymod0}, we show how the misalignment angle varies as a function of~$f_{\rm L}$~(see colors) for both hemispheres~(see diagonal perpendicular hatches) given $\delta l_i = \pm 12^\circ$ in the southern and northern hemispheres, respectively. With light-gray shades, we also show the case with $\delta l_i = +12^\circ$ in both hemispheres.

\begin{figure}[!t]
\begin{center}
\resizebox{0.8\hsize}{!}{\includegraphics{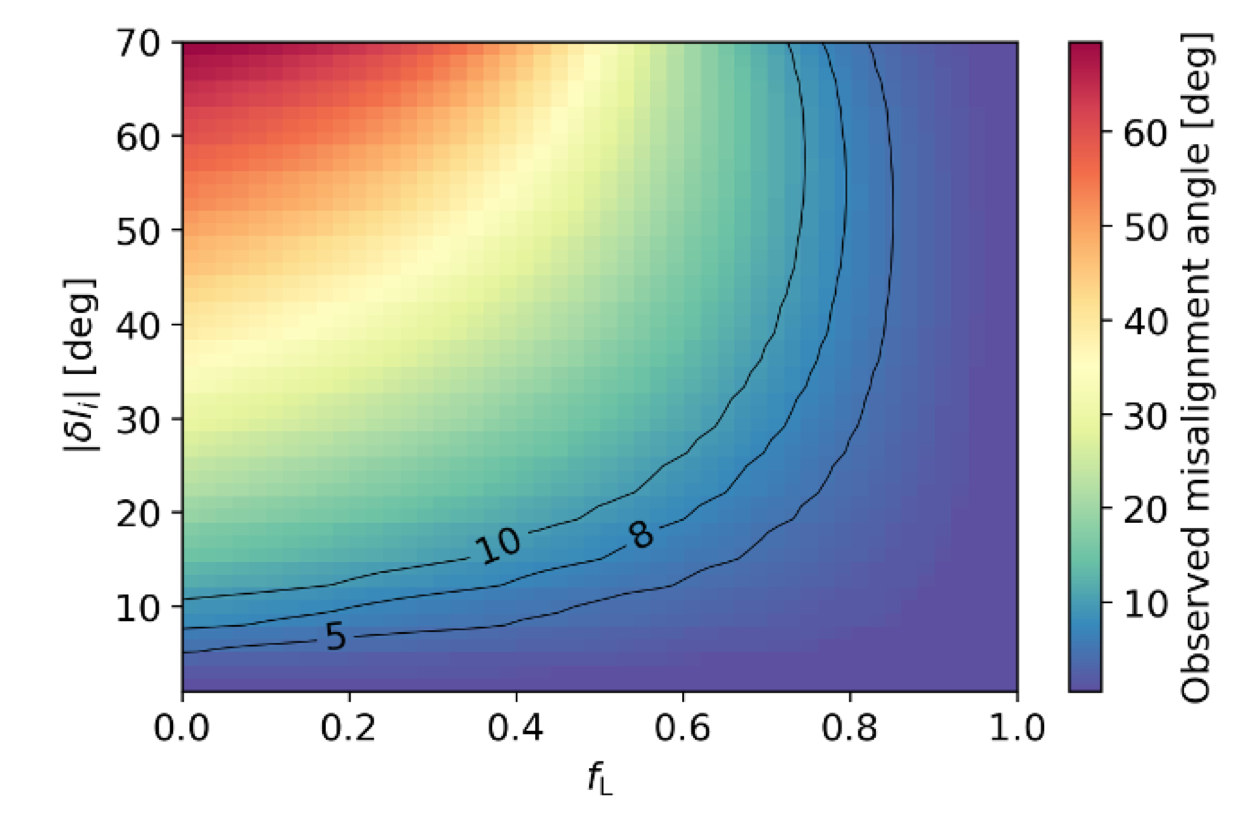}}
\caption{Dependence of the observed misalignment angle with~$f_{\rm L}$ and~$|\delta l_i|$ in the toy model. Black contours show levels of misalignment angle similar to what was measured in the \Planck\ data.}
\label{fig:toymod1}
\end{center}
\end{figure}

We observe that only tilted distortion of the LB field compared to the regular field~(see \figref{fig:sketch}) or a change in sign of~$\delta l_i$ between the two Galactic hemispheres can reproduce the same hemispherical sign of misalignment angle observed in the data. This may correspond to the distortion of the regular field caused by the LB similar to an effective large-scale helical component of the magnetic field, as proposed by \citet{Bracco2019a}.   
We also note that, despite having $\delta l_i = \pm 12^\circ$, the observed misalignment angle can be significantly smaller depending on~$f_{\rm L}$. Specifically, it ranges from~$0^\circ$ to~$12^\circ$ depending on whether the LB completely dominates the total polarization signal~($f_{\rm L} =1$) or is negligible compared to the regular-field contribution~($f_{\rm L} =0$). In \figref{fig:toymod1}, we further explore this by varying both~$f_{\rm L}$ and~$|\delta l_i|$ and applying opposite signs in the two hemispheres. Our models demonstrate that even with large values of~$|\delta l_i|$ (e.g., $|\delta l_i| > 50^\circ$) the observed misalignment may be negligible depending on~$f_{\rm L}$. With black contours we draw the levels of misalignment observed in the data between~$5^\circ$ and~$10^\circ$.

\subsubsection{Synchrotron misalignment}\label{ssec:sync}

These models predict that the misalignment with the \HI~templates should increase for datasets with less contribution from the LB, i.e., when $f_{\rm L}$~is smaller than for dust polarization.
This hypothesis is supported by \Planck\ polarization data at 30~GHz, which predominantly traces synchrotron radiation. Given that the synchrotron scale height in spiral galaxies is generally larger than that of the dusty disk over scales of a few hundred parsecs~\citep[e.g.,][]{Beck2015}, synchrotron radiation is expected to have a smaller~$f_{\rm L}$ compared to the dusty case, that is, the contribution of the regular magnetic field should be more important~\citep{Pelgrims2025}. In \figref{fig:Dpsi_scale_30}, the NPDFs of~$\Delta \psi$ using \Planck\ data at 30~GHz are shown with the respective values of~$\langle\Delta \psi\rangle_{\rm circ}$ as a function of~$\ell_{\rm FWHM}$. Both hemispheres exhibit the same sign of large-scale misalignment as observed with dust, but with significantly larger values of~$\langle\Delta \psi\rangle_{\rm circ}$ aligning with the model expectations. Furthermore, the scale dependence of~$\langle\Delta \psi\rangle_{\rm circ}$ in the two hemispheres, similar to that observed for dust, is particularly notable in the case of synchrotron radiation. While there is little dependence on~$\ell_{\rm FWHM}$ in the northern hemisphere, an abrupt change in misalignment is observed in the southern hemisphere. These features highlight the multi-scale complexity of the misalignment, which our simplified model does not fully capture. A more detailed comparison of synchrotron and dust polarization would be important but is beyond the scope of this work.
    
\subsubsection{Starlight polarization: Distance dependence}\label{sssec:stars}
\begin{figure*}[!h]
\begin{center}
\resizebox{0.85\hsize}{!}{\includegraphics{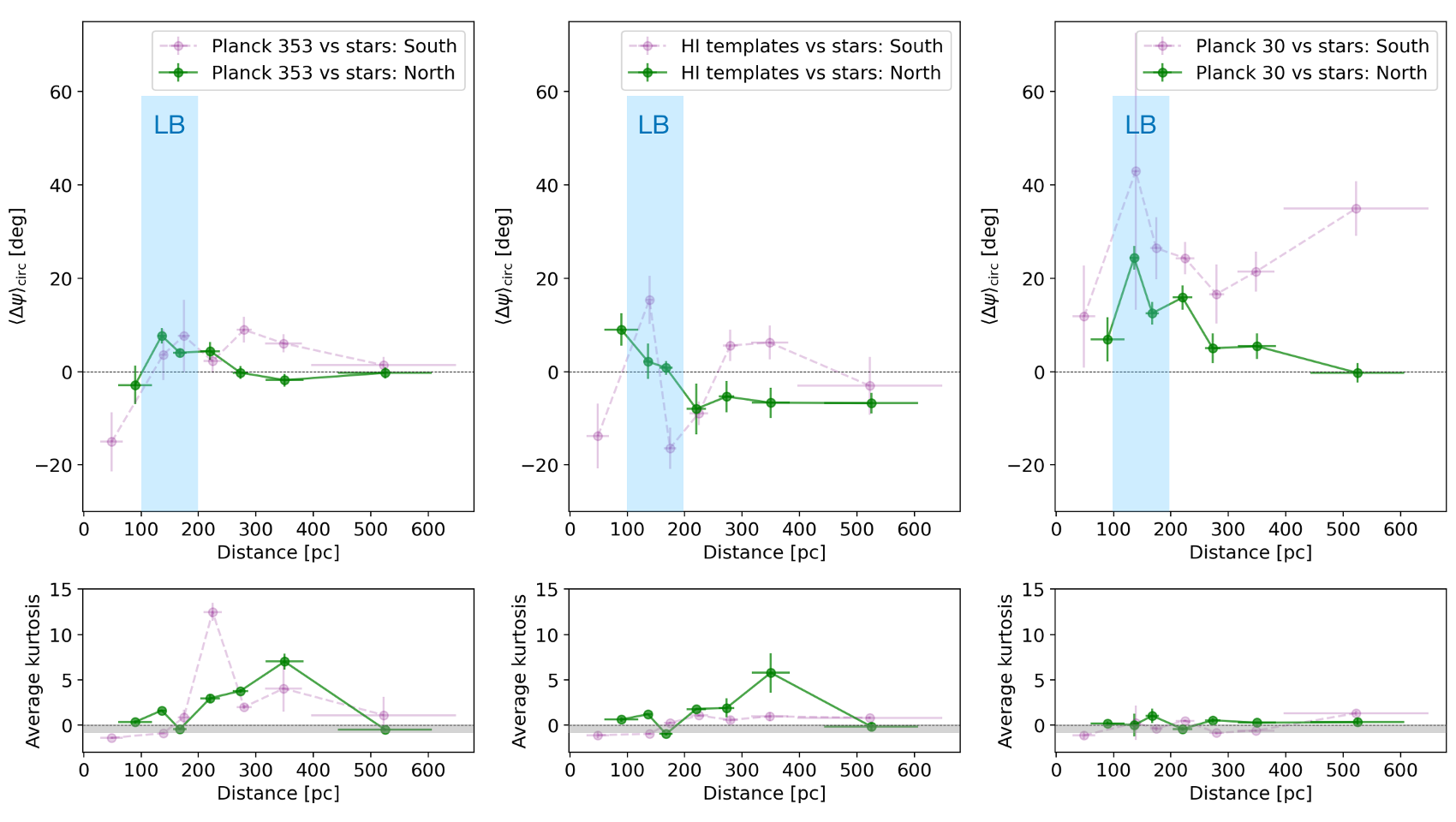}}
\caption{Misalignment angle between polarization data~(rotated by~90$^{\circ}$) and starlight polarization measurements as a function of the stellar distance for \Planck\ data at 353~GHz~(top-left panel), \HI~polarization templates~(top-central panel), and \Planck\ data at 30~GHz~(top-right panel). The misalignment is shown in green and pink for the northern and southern hemispheres, respectively. The misalignment peaks at the edge of the LB in cyan. With the same color scheme, the bottom panels show the average normalized kurtosis of the corresponding $\Delta \psi$~distributions. The uncertainty is the standard deviation between the two versions of each dataset. The gray shaded area represents the standard deviation of the average kurtosis of a normal distribution in the case of low-number statistics, as for the number of stars per distance bin~(approximately 20).} 
\label{fig:miso_stars}
\end{center}
\end{figure*}
The geometrical toy model suggests that the misalignment angle should also depend on the heliocentric distance to the magnetic field responsible for the dust polarization signal. We explored this idea using the starlight polarization catalog described in \secref{ssec:stars}.
We computed the misalignment angle between starlight polarization angles and the \Planck\ data at 353~GHz, the \HI~templates, and the \Planck\ data at 30~GHz. This was obtained in the 20\%~mask by sampling the starlight measurements in seven distance bins within 1~kpc from the Sun. The total number of selected stars is~306, corresponding to 86\% of the original catalogs of \citet{Heiles2000}, \citet{Berdyugin2001}, \citet{Berdyugin2002}, and \citet{Berdyugin2014}.

We applied \eqref{eq:Dpsi} after projecting the starlight measurements on a {\tt{HEALPix}} grid at $N_{\rm side} = 128$. Before computing the misalignment angle, we averaged the Stokes parameters of those stars with Galactic coordinates falling in the same {\tt{HEALPix}} pixels. However, given the sparsity of stars at high Galactic latitudes~(see \figref{fig:star_hits}), this averaging effect did not strongly impact the results, as it involved multiple counting for only 3\% of the sample. We note that stellar sparsity also represents a potential systematic bias in estimating large-scale polarization fields using stars, as some coherence from small to large scales must be assumed.

We computed the misalignment angle for the northern and southern hemispheres, separately. The number of bins was chosen to have a roughly homogeneous number of stars per bin while also sampling the distance range between a few tens to hundreds of parsecs. The upper limits of each bin were~120, 150, 190, 250, 300, 420, and 1000~pc. With 208 and 98~stars in the northern and southern Galactic hemispheres, respectively, seven distance bins guaranteed, on average, 30 and 14~stars per bin in the two hemispheres. 

In \figref{fig:miso_stars}, we show the misalignment angle between starlight polarization and the other tracers as a function of the average stellar distance per bin. Error bars on the $y$~axis represent systematic effects of \Planck\ and \HI~data and the dispersion of the starlight polarization angles. On the $x$~axis, the error bars indicate the distance width of each bin. In the bottom row of the same figure, we display the kurtosis of the corresponding NPDFs averaged between two versions of the same dataset~(e.g., PR4 and {\tt{SRoll2}} in the case of \Planck\ data at 353~GHz). The normalized kurtosis provide us with an indication on how much the NPDFs are peaked. The gray-shaded areas represent the standard deviation of the average kurtosis of a normal distribution. 
Apart from the first distance bin, all NPDFs are well-peaked around the value of~$\langle\Delta \psi\rangle_{\rm circ}$ plotted in the top row of \figref{fig:miso_stars}. 

We observe that, at 353~GHz, a misalignment angle on the order of~$7^\circ$, which is consistent with \figref{fig:Dpsi_masks}, is found at the distance of the LB wall, namely, between 100 and 200~pc at high Galactic latitude~\citep[e.g.,][]{Pelgrims2020,Oneill2024}. The misalignment disappears at larger distances, in agreement with~\citet{Skalidis2019}. Probably because of better number statistics, this effect is clearer in the northern hemisphere than in the southern hemisphere. However, this could also be a physical effect as illustrated by the hemispherical difference in Figs.~\ref{fig:Dpsi_masks} and~\ref{fig:Dpsi_scale}. 

As predicted by the geometrical model in \secref{ssec:modelproj}, starlight polarization and \Planck\ data at 353~GHz are sensitive to the same magnetic-field structure only at large heliocentric distances; at smaller distances, stars are sensitive to the LB/CNM field revealed by the misalignment. This is supported by the misalignment angle between stars and the \HI~templates. Starlight polarization angles show no misalignment with the \HI~templates at the distance of the LB wall but reveal a flat misalignment at further distances, consistent with dust polarization in emission. 

Finally, we observe a stronger positive misalignment between starlight polarization and \Planck\ data at 30~GHz in both hemispheres. This misalignment vanishes more slowly with distance compared to the \Planck\ data at 353~GHz in the northern hemisphere, and it remains roughly constant in the southern hemisphere. Understanding these differences in detail is beyond the scope of the present paper; however we speculate that they may be related to a generally more coherent magnetic-field structure along the line of sight in the northern area sampled by the stellar measurements. These results are consistent with the proposed scenario, where the 30-GHz data would more efficiently trace the regular magnetic-field component compared to the LB field, implying a smaller value of~$f_{\rm L}$, thus a stronger misalignment angle.
In summary, starlight polarization data support our phase superposition scenario as follows:
\begin{itemize}
    \item At the location of the LB, stars mostly trace CNM and its magnetic field while dust emission also includes the WNM-weighted field on larger scales, producing misalignment. At large distances, stars additionally probe the WNM-weighted magnetic field and begin converging to the dust emission.
    \item Complementary to the previous case, \HI\ templates correlate with starlight polarization at the location of the LB and depart from it at larger distances. 
    \item The typical scale height is larger for synchrotron than for dust. Because synchrotron mostly traces the mean field on scales larger than the LB (low value of $f_{\rm L}$), it shows a strong misalignment with stars at the position of the LB and only correlates with starlight polarization at very large distances in the northern hemisphere, where the magnetic-field structure may be generally more coherent in the area sampled by the stars.  
\end{itemize}

A denser sample of stars, with defined distances and polarization measurements, will be essential for a better investigation of our scenario regarding the misalignment angle. Projects such as the Polar-Areas Stellar Imaging in Polarization High-Accuracy Experiment~\citep[PASIPHAE,][]{Tassis2018}, which will increase the number of studied stars a thousandfold over the current state of the art at intermediate and high Galactic latitudes, are expected to be transformative.

\subsection{Impact on dust polarization angular power spectra}\label{sec:powerspectra}

We now discuss the link between the observed large-scale misalignment and the dust polarization angular power spectra, particularly focusing on the observed $TB$~correlation in the \Planck\ data~\citep{PlanckXI2020}. We first present the \Planck\ $TB$~power spectra at 353~GHz for the two Galactic hemispheres and then demonstrate how the signal can be suppressed by removing the large-scale contribution~(\secref{ssec:powerspectra_data}). 

In its current form, our model (see Sect.~\ref{ssec:modelproj}) does not reproduce $TB$ cross-spectra. Future work is needed in order to include multiscale filamentary intensity models that must correlate with the 3D magnetic-field structure as in the geometrical model described in this work and similar to the case study of line-of-sight superposition presented in \citet{Vacher2023}. Another possibility could be to look at realistic MHD simulations in projections, as initiated in previous works \citep[e.g.,][]{Clark2021,Pelgrims2022,Maconi2023}. Nevertheless, in \secref{ssec:powerspectra_model}, we use synthetic filamentary models in 2D, as described in \appref{app:filamod}, to illustrate the effect of a coherent large-scale misalignment on the small-scale $TB$~correlation.   

\subsubsection{Hemispherical look at the \Planck\  data}\label{ssec:powerspectra_data}

\begin{figure}[!h]
\begin{center}
\resizebox{1\hsize}{!}{\includegraphics{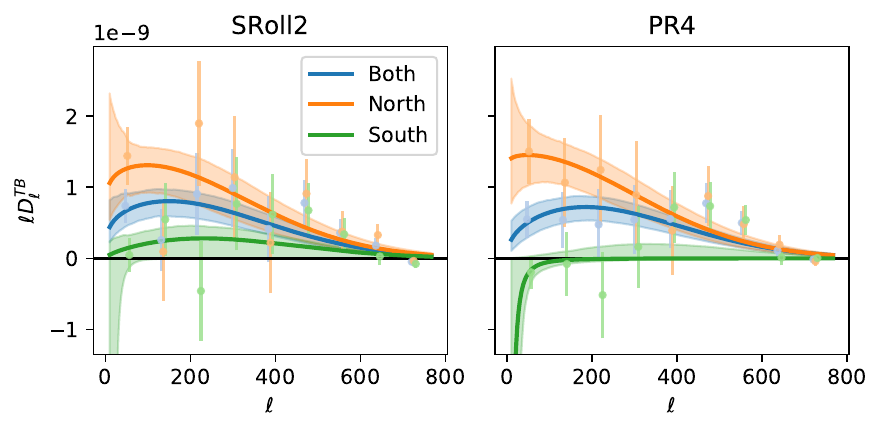}}
\caption{\Planck\ $TB$~power spectra in the northern and southern Galactic hemispheres. Power spectra were computed in the 20\%~mask with both {\tt{SRoll2}} and PR4 maps. Units are in~$K^2_{\rm CMB}$.} 
\label{fig:dataspec}
\end{center}
\end{figure}

In \figref{fig:dataspec}, we show the $TB$~power spectra, defined as $\mathcal{D}^{TB}{\ell} = \ell(\ell+1)C^{TB}{\ell}/2\pi$ and computed in the 20\%~mask for the northern and southern Galactic hemispheres. The details of the power-spectrum calculation are provided in \appref{app:TB}. Using both {\tt SRoll2}~(left panel) and PR4 data~(right panel), the spectra are averaged with a linear multipole binning starting at $\ell = 5$~(shown as data points). Best-fit power-laws with a beam roll-off are also displayed, with shaded regions indicating the 1$\sigma$~uncertainty. The northern hemisphere consistently exhibits a stronger positive $TB$~correlation than the southern hemisphere.

A similar hemispherical split was performed in \cite{Cukierman2023}, which focused mainly on higher multipoles~($\ell > 100$) and larger sky areas~(e.g., $f_\mathrm{sky} = 70\%$). In fig.~11, \citet{Cukierman2023} provides a misalignment estimate based on~$T_\mathrm{d} B_\mathrm{d}$~(labeled ``Dust only'') for the 20\%~mask and the associated hemispherical splits, though the scales have been limited to $\ell > 100$. As in Fig.~\ref{fig:dataspec}, the north yields a signal that is moderately strong, and the south yields a result that is consistent with zero.

\begin{figure}[!t]
\begin{center}
\resizebox{0.8\hsize}{!}{\includegraphics{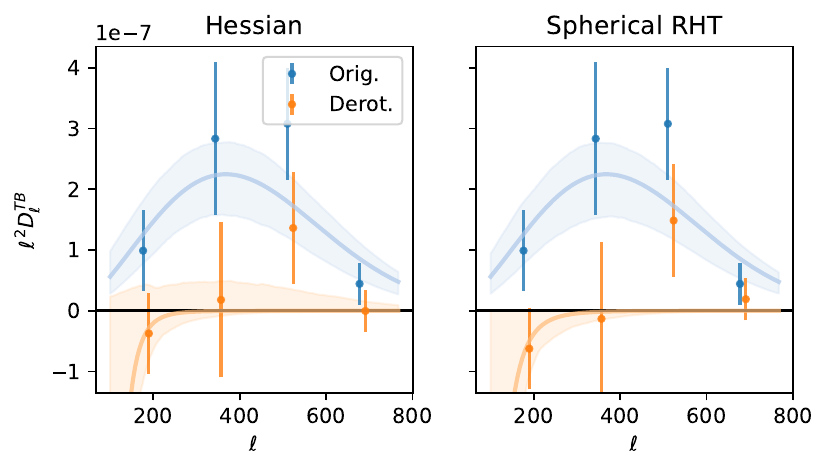}}
\caption{\Planck\ $TB$~power spectra from both Galactic hemispheres with the 20\%~mask before and after de-rotation, using {\tt SRoll2} for the dust maps and two different \HI~templates. The de-rotation was computed at $\ell_{\rm ref} = 20$. Units are~$K^2_{\rm CMB}$.} 
\label{fig:dataspec_derot}
\end{center}
\end{figure}

This hemispherical difference is consistent with the expectations from the misalignment analysis, where a stronger and more coherent large-scale misalignment angle is found in the north~(see also Fig. \ref{fig:TBHI_spectra}).
To further establish the connection, we recomputed the spectra after de-rotating the large-scale misalignment, which is measured from the angle differences between smoothed \Planck\ polarization maps and smoothed \HI~templates. This calculation uses Eqs.~\ref{eq:Dpsi} and~\ref{eq:rot} with $\ell_{\rm ref} = 20$. In \figref{fig:dataspec_derot}, we show the original and de-rotated power spectra; for the latter, we consider both \HI~templates, namely, the Hessian and the SRHT. In both cases, we are able to suppress the $TB$~correlation in the north by accounting only for the large-scale dust-\HI\ misalignment. We stress that the de-rotation, which is based only on large-scale information~($\ell_{\rm ref} = 20$), is able to suppress the $TB$~correlation even at small scales~($\ell > 200$). 
Finally, we notice that hemispherical differences are also found with other tracers of the multiphase and magnetized ISM, such as rotation-measure patterns at large angular scales, whose origin remains unclear~\citep[e.g.,][]{Dickey2022, Booth2025}. 

\subsubsection{Insights from 2D synthetic filamentary models}\label{ssec:powerspectra_model}

As described in \appref{app:filamod}, we built synthetic maps to test the impact of a large-scale misalignment on the $TB$~correlation at small scales. The synthetic maps include a filamentary model perfectly aligned to the magnetic field orientation, to which we add two uniform effective misalignment angles projected on the plane of the sky, $\Delta \psi_{\rm N}$ and~$\Delta \psi_{\rm S}$, in the two Galactic hemispheres. These two components correspond to the LB magnetic field and the total one as described by the geometrical model in \secref{ssec:modelproj}. From the synthetic Stokes~$IQU$, we compute~$\ell^2{D}^{TB}_{\ell}$ in arbitrary units within the 20\%~mask using the {\tt{healpy}} package. Rather than fitting the data, our goal is to gain intuition on the effects of the geometrical model on the polarization power spectra.   
\begin{figure}[t]
\begin{center}
\resizebox{0.8\hsize}{!}{\includegraphics{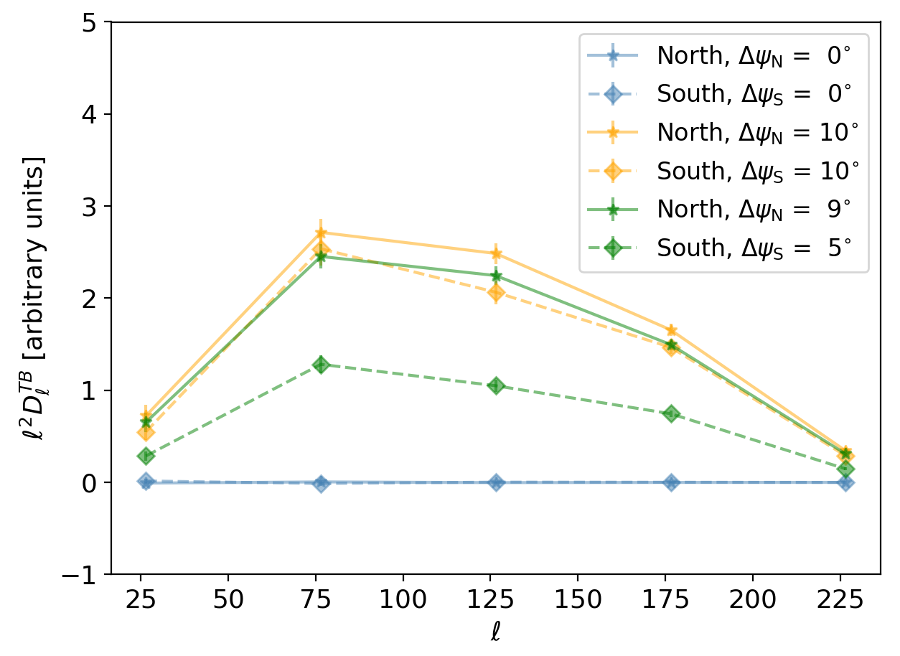}}
\caption{Modeled $TB$~power spectra as a function of the misalignment angles~$\Delta \psi_{\rm N}$ and~$\Delta \psi_{\rm S}$ in the north and south Galactic hemispheres, respectively. Power spectra were computed with the 20\%~mask.} 
\label{fig:modSpec}
\end{center}
\end{figure}
As shown in \figref{fig:modSpec}, by setting different values of~$\Delta \psi_{\rm N}$ and~$\Delta \psi_{\rm S}$ at the largest scales~(uniform over each Galactic hemisphere), we can control the level of $TB$~correlation over a wide range of multipoles. Only by setting the same sign for~$\Delta \psi_{\rm N}$ and~$\Delta \psi_{\rm S}$\footnote{Given the plane-of-the-sky projection and the 3D modeling, the same sign between $\Delta\psi_{\rm N}$ and $\Delta\psi_{\rm S}$ corresponds to opposite signs in the case of $\delta l_i$.} can we achieve the same sign of~$\ell^2{D}^{TB}_{\ell}$ in the two hemispheres. This effective tilt of the LB field could be analogous to the large-scale helical magnetic field proposed by \citet{Bracco2019a} to reproduce the observed $TB$~signal.

\section{Summary and conclusion}\label{sec:sum}

We have presented new data analyses and one physical scenario to gain insight into the origin of two observables of the dusty, polarized sky at intermediate and high Galactic latitudes: 1)~the statistical misalignment between atomic hydrogen~(\HI) filamentary structures and dust polarization angles and 2)~the $TB$~correlation detected in \Planck\ polarization data. Using a multiwavelength analysis of \Planck\ data at 30, 217, and 353~GHz as well as starlight polarization measurements and \HI-based templates, we showed that both phenomena are stronger at large scale and in the northern Galactic hemisphere. The misalignment could be interpreted as a consequence of large-scale line-of-sight projection effects on the magnetic-field structure sampled by different ISM phases in the solar neighborhood. In power spectra, this large angular scale effect could be inherited by small scales, which have been considered responsible for both observables~\citep[e.g.,][]{Huffenberger2020, HerviasCampos2022, Clark2021, Cukierman2023, HerviasCampos2025}. The key results of this work are the following:
\begin{itemize}
 \item The misalignment angle in the 20\%~of the sky at high Galactic latitude is significant mainly on large angular scales~($\ell \leq 20$) and varies between the two Galactic hemispheres, with the northern hemisphere showing a larger and more coherent misalignment. This result holds true at all \Planck\ frequencies considered in this work, being stronger at lower frequencies in the synchrotron domain. 
 
 \item We showed that the misalignment is unlikely to be related to dust emissivity variations and can be reproduced with a two-layer geometrical toy model where the large-scale regular magnetic field in the solar neighborhood is distorted by the LB. The LB induces a relative tilt between CNM- and WNM-traced magnetic fields, which in projection produces the misalignment and, possibly, the $TB$~correlation. The observed distance dependence of the misalignment, traced with starlight polarization measurements, supports this interpretation for the misalignment. 

 \item As a consequence of the proposed scenario, \HI~filamentary structures can be considered statistically aligned with magnetic fields in the diffuse ISM, in agreement with hydro-magnetic turbulence, although some caution is needed. The projected scatter around the mean remains significant, on the order of a few tens of degrees. Future work will be necessary to clarify the physical origin of this scatter. 

 \item We presented $TB$~power spectra of \Planck\ data at 353~GHz for the two Galactic hemispheres. We found that the $TB$~correlation is stronger at large scales and mostly in the northern hemisphere, consistent with the misalignment analysis. By de-rotating the Stokes parameters to account for the large-scale misalignment between the \Planck\ data and \HI~templates~($\ell \leq 20$), we were able to consistently suppress the observed $TB$~correlation at small scales~($\ell > 100$). This de-rotation is more effective in the northern hemisphere than in the southern, where the scale dependence of the misalignment angle suggests a more complex physical scenario. 
 
 \item Although we could not reproduce the complexity of the $TB$ correlation with the toy model detailed in Sect.~\ref{ssec:modelproj}, using synthetic 2D filamentary sky models we demonstrated that a large-scale misalignment can produce a small-scale $TB$~correlation. 
\end{itemize}

Our results emphasize the critical role of large-scale structures in the solar neighborhood in shaping polarized Galactic signals. This has important implications for both Galactic magnetic-field studies and future CMB polarization experiments.

\begin{acknowledgements}  
The authors acknowledge the Interstellar Institute’s programs “II6” \& “II7” and the Paris-Saclay University’s Institut Pascal for hosting discussions that nourished the development of the ideas behind this work. This work was supported by NSF grant AST-2109127. A.B. acknowledges financial support from the INAF initiative "IAF Astronomy Fellowships in Italy" (grant name MEGASKAT). This work is a tribute to the imagination of my beloved uncle Pasquale: {\it "So be it, heart; bid farewell without end" [H. Hesse]}. R.S. was supported by NASA through the NASA Hubble Fellowship grant HST-HF2-51566.001 awarded by the Space Telescope Science Institute, which is operated by the Association of Universities for Research in Astronomy, Inc., for NASA, under contract NAS5-26555. We are thankful to Vincent Pelgrims, Susan Clark, Antoine Marchal, Pierre Lesaffre, and Tuhin Ghosh for insightful discussions. We thank Matthew A. Price for helping us with the {\tt{S2WAV}} package. Some of the results in this paper have been derived using the {\tt healpy} and {\tt HEALPix} packages. In the analysis we made use of {\tt astropy} \citep{astropy2018}, {\tt scipy} \citep{Virtanen2020}, and {\tt numpy} \citep{Harris2020}.
\end{acknowledgements}

\bibliographystyle{aa}
\bibliography{biblio}

\appendix
\onecolumn
\section{Sky maps}\label{app:figures}

\begin{figure}[h]
\begin{center}
\resizebox{1\hsize}{!}{\includegraphics{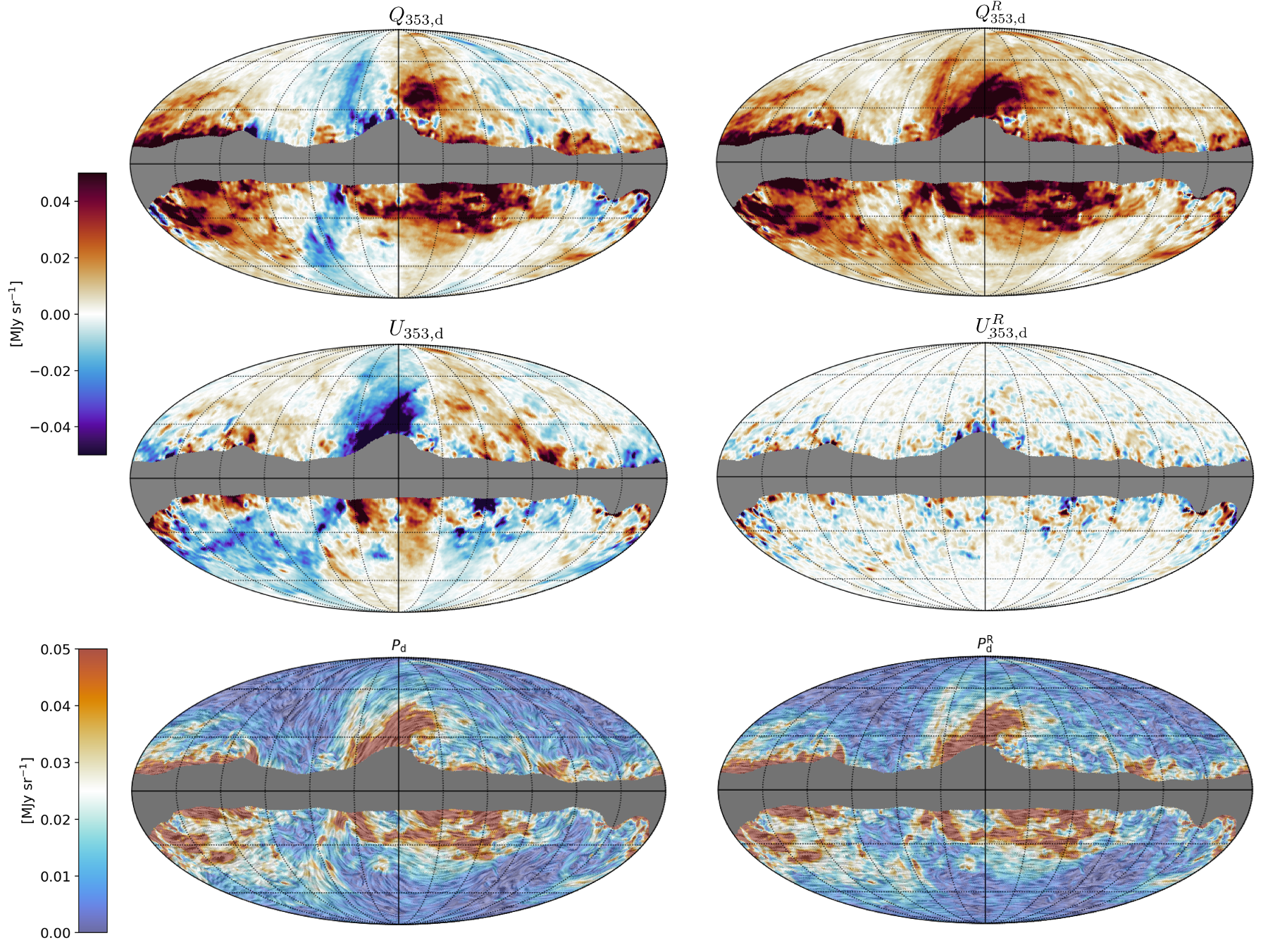}}
\caption{Effect of the angle rotation on the \Planck\ Stokes parameters using \eqref{eq:rot}. The original data at FWHM=80$\arcmin$ are shown in the left column and the maps rotated with respect to $\ell_{\rm ref} = 20$ in the right column. The Stokes parameters are shown in the top and central panels with the same color bar displayed on the left, while polarized intensities are shown in the bottom panels in the background of the drapery patterns tracing the magnetic-field orientation obtained through LIC. The 80\%~sky mask is applied~(see \figref{fig:masks}).} 
\label{fig:stokes}
\end{center}
\end{figure}

As described in \secref{sec:methods}, \figref{fig:stokes} shows an example of rotating the Stokes parameters with respect to a larger-scale reference set by $\ell_{\rm ref} =20$. The rotation obtained with \eqref{eq:rot} is applied on~$Q_{\rm 353, d}$ and~$U_{\rm 353, d}$ from {\tt{SRoll2}}. We applied the 80\%~mask. After rotation, it can be noticed that most of the polarization signal is converted, by construction, to positive values of~$Q^{\rm R}_{\rm 353, d}$, while $U^{\rm R}_{\rm 353, d}$~resembles a dispersion around the mean. In this reference frame, the polarization angle, shown with the line integral convolution~(LIC) function of {\tt{healpy}} in the bottom row, is essentially perpendicular to the Galactic plane and the magnetic field orientation parallel to it.

\section{Misalignment angle at 217 and 30 GHz}\label{app:otherfreq}

\begin{figure*}[!h]
\begin{center}
\resizebox{.9\hsize}{!}{\includegraphics{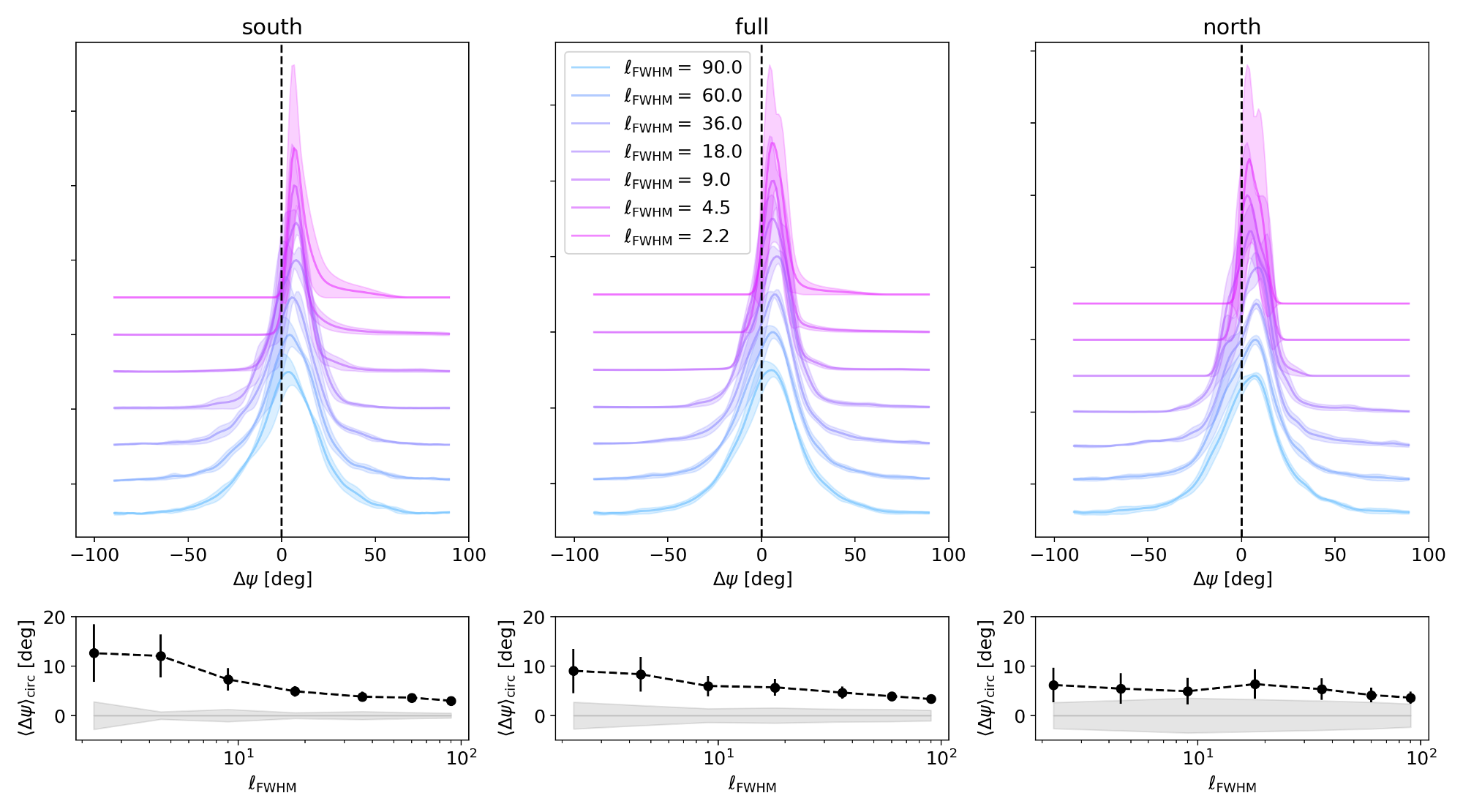}}
\caption{Same as for \figref{fig:Dpsi_scale} but considering \Planck\ data at 217~GHz as a test of the robustness of the results for 353~GHz. A similar misalignment as at 353 GHz is also found at 217 GHz.} 
\label{fig:Dpsi_scale_217}
\end{center}
\end{figure*}

\begin{figure*}[!h]
\begin{center}
\resizebox{0.9\hsize}{!}{\includegraphics{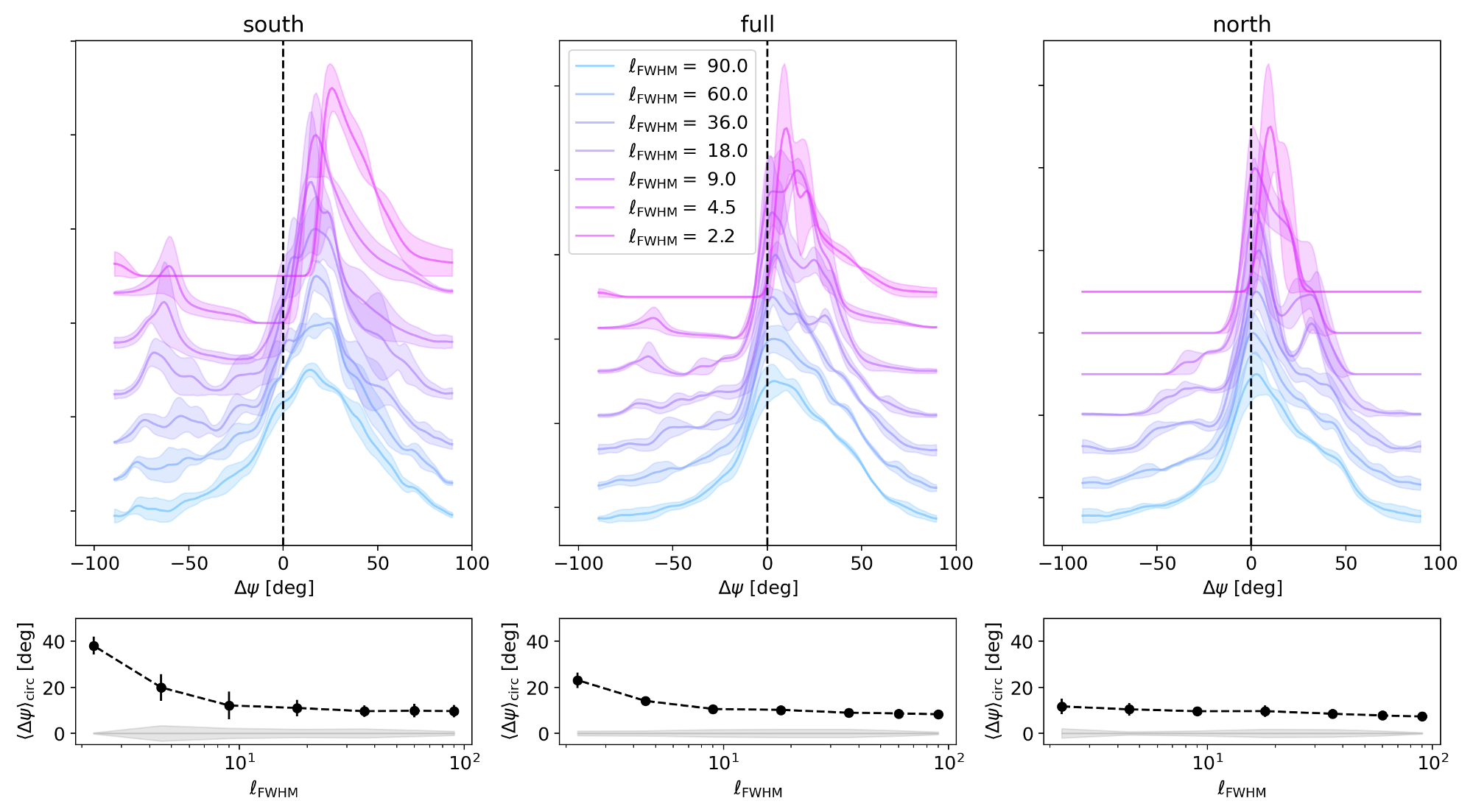}}
\caption{Same as for \figref{fig:Dpsi_scale} but considering \Planck\ data at 30~GHz, which is dominated by synchrotron radiation rather than dust emission. A stronger misalignment than at 353 GHz is found at 30 GHz, where synchrotron polarization is traced.} 
\label{fig:Dpsi_scale_30}
\end{center}
\end{figure*}

In this appendix, we show the NPDFs of the misalignment angle as a function of~$\ell_{\rm FWHM}$ between the \HI~templates and the \Planck\ data at~217 and 30~GHz, respectively. The former is shown in \figref{fig:Dpsi_scale_217}, and the latter in \figref{fig:Dpsi_scale_30}. In both cases we found a strong large-scale misalignment as in the case of \Planck\ data at 353 GHz. Refer to \secref{ssec:dust} and \secref{ssec:modelproj} for more details. 

\section{Starlight measurements on the sky}\label{app:starhits}

\begin{figure*}[!h]
\begin{center}
\resizebox{0.7\hsize}{!}{\includegraphics{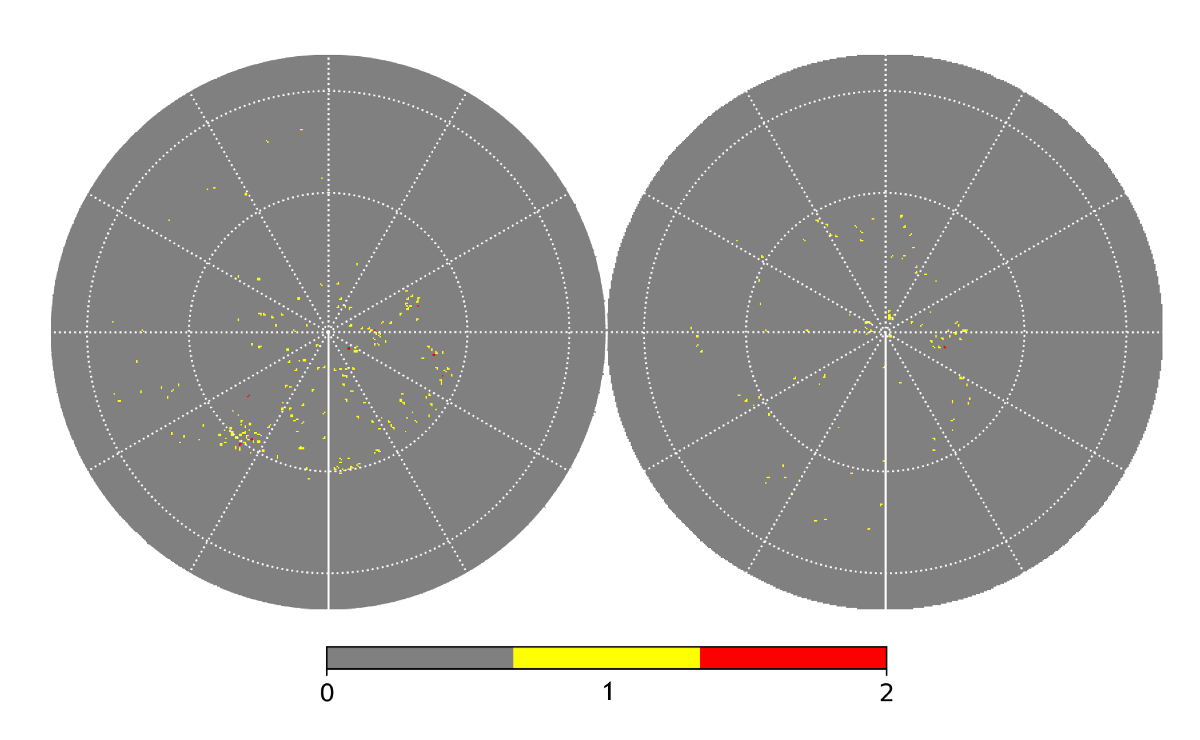}}
\caption{Orthographic projection around the northern~(left) and southern~(right) Galactic poles of the hit map of stars after reprojecting the measurements on a {\tt{HEALPix}} grid at $N_{\rm side} = 128$. A galactic coordinate grid is overlaid.} 
\label{fig:star_hits}
\end{center}
\end{figure*}

This appendix shows the distribution of starlight measurements in the 20\%~mask with an orthographic projection around the two Galactic poles. In \figref{fig:star_hits}, the star hits are displayed, with only 3\%~of pixels counting two selected stars in the stellar catalog at $N_{\rm side} =128$.

\section{Computing $TB$ power spectra}\label{app:TB}

Angular power spectra~$C_{\ell}$ are computed with {\tt NaMaster}~\citep{Alonso2019}. Our sky masks are apodized with a $C^2$~window~\citep{Grain2009} with a scale of~$1^\circ$. 
For the computation of power spectra, different from what is described in \secref{sec:data}, all of our maps are smoothed to~$40\arcmin$ and downgraded to $N_{\rm side}=256$ in {\tt HEALPix} format~\citep{Gorski2005}. 
Our power spectra are computed in the $TEB$~basis~\citep[e.g.,][]{Zaldarriaga1997}, and we are mainly interested in the $TB$~cross spectrum.

\begin{figure*}[!h]
\begin{center}
\resizebox{.8\hsize}{!}{\includegraphics{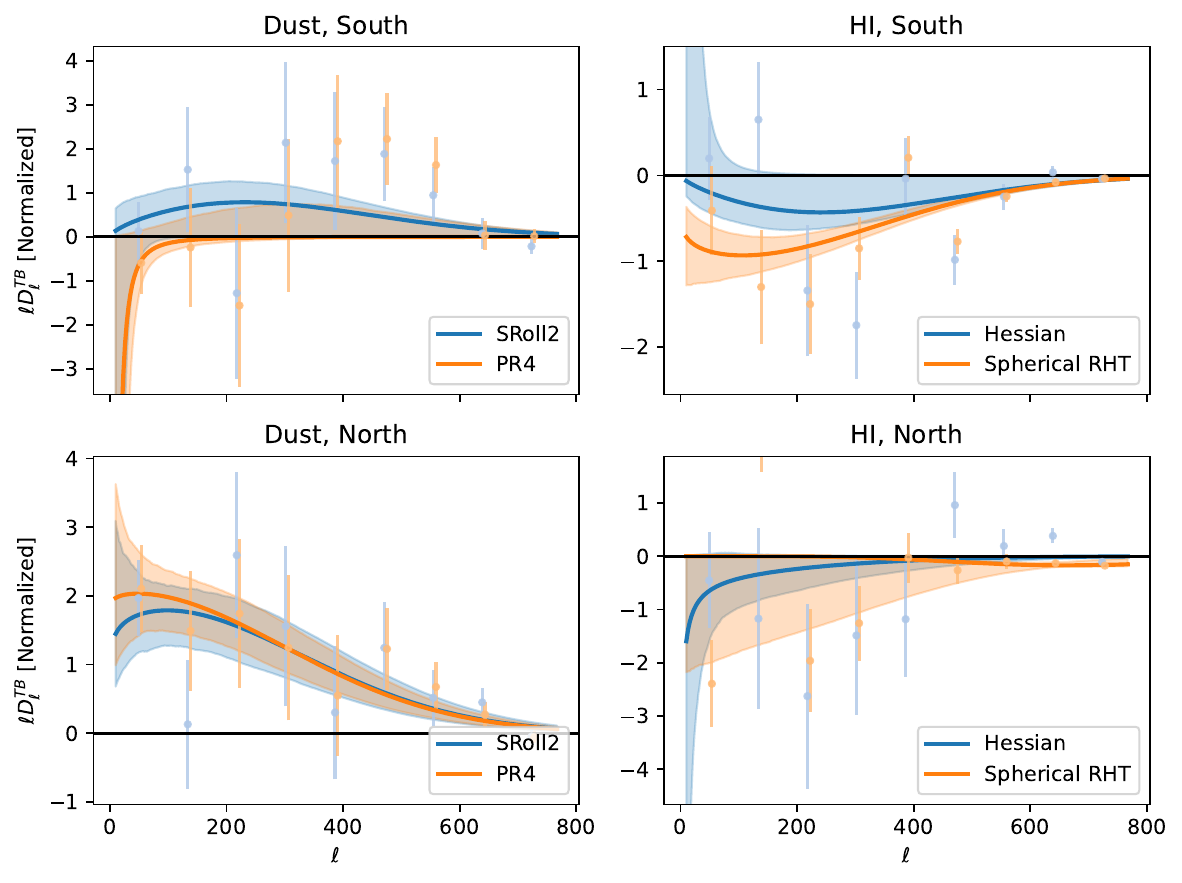}}
\caption{Estimates of $TB$~cross spectra for dust and \HI~filaments in the two Galactic hemispheres with the 20\%~mask.  This is a comparison between the various spectra depending on the given \Planck\ dataset~(left, $T_{\rm d} B_{353}$) and \HI~template~(right, $T_{\rm d} B_{\rm HI}$). To be able to compare all of these quantities, which have different physical units, we have normalized each spectrum to the mean of its absolute values.} 
\label{fig:TBHI_spectra}
\end{center}
\end{figure*}

In \figref{fig:TBHI_spectra}, we present the $TB$~power spectra for four distinct cases in the form of $D_{\ell} = \ell (\ell +1)C_{\ell}/2\pi$. The analysis separates the two Galactic hemispheres, and the input $QU$~Stokes parameters are derived either from \Planck\ data at 353~GHz or from the \HI~templates. For the intensity map, we consider only~$I_{\rm d}$~(hereafter $T_{\rm d}$, see \secref{ssec:planck}). To assess the impact of systematic effects, we compute the power spectra using two versions of each dataset~(e.g., {\tt{SRoll2}} and PR4 for the \Planck\ data). For ease of comparison, each spectrum is normalized to the mean of the absolute values of its bandpowers. A significant positive correlation in the dust $TB$~power spectra is observed only in the northern hemisphere, consistent across both the {\tt{SRoll2}} and PR4 maps. 

When the $B$~modes are drawn from the \HI~templates, we find a hint of $T_\mathrm{d} B_{\mathrm{HI}} < 0$, but this is significant only in the southern hemisphere and only for the spherical RHT. Since the \HI~templates are constructed entirely independently of dust polarization data, a true signal of $T_\mathrm{d}B_\mathrm{HI} < 0$ would indicate that the \HI~filamentary morphology displays a chiral asymmetry. This possibility was noted in the conclusion and appendix of \citet{Cukierman2023}. A morphological chirality would present an additional contribution to the dust~$TB$. If the morphological contribution is negative, it may partially cancel the contribution from magnetic misalignment. In the absence of a strongly nonzero $T_\mathrm{d}B_\mathrm{HI}$~signal, we consider the misalignment effect to be the main contributor to~$T_\mathrm{d}B_\mathrm{d}$.

\section{Synthetic filamentary all-sky model}\label{app:filamod}

\begin{figure*}[!t]
\begin{center}
\resizebox{1\hsize}{!}{\includegraphics{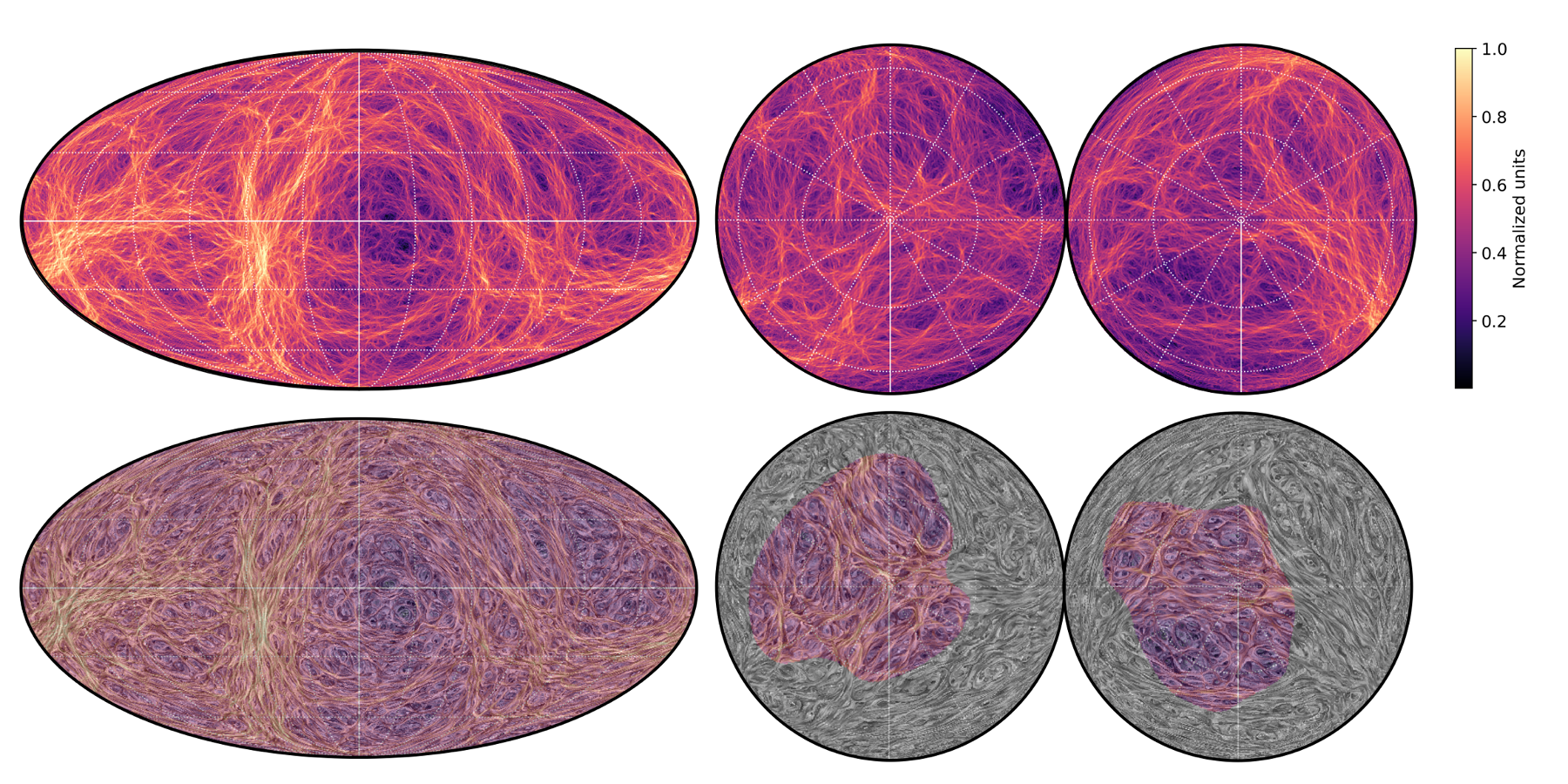}}
\caption{Mollweide~(left) and orthographic~(right) projections of one synthetic filamentary model with misalignment around the poles. In colors in the top, we show the total intensity model in normalized units, while in the bottom overlaid drapery patterns also trace the modeled magnetic-field orientation. The orthographic projections are centered around the Galactic poles. In the bottom, we apply the 20\%~sky mask. The misalignment is uniform and is fixed to~$9^\circ$ in the north and~$5^\circ$ in the south.} 
\label{fig:modFil}
\end{center}
\end{figure*}

This appendix describes the procedure to create all-sky models of non-Gaussian filamentary structures from realizations of synthetic pseudo-random fields, as shown in \figref{fig:modFil}. Our approach extends to {\tt{HEALPix}} spherical grids the {\tt{pywavan}}\footnote{\url{http://github.com/jfrob27/pywavan}} technique developed in Python by \citet{Robitaille2020} for flat-sky models. 
The key principle of \citet{Robitaille2020} is building a statistical model based on multiplicative random cascades, which are designed to replicate the multi-fractal, hierarchical structure of intermittent features developed in turbulent media such as the ISM. They presented a version of the multiplicative process, where the spatial fluctuations as a function of scale are produced with wavelet transforms of fractional Brownian motion~(FBM) realizations. Using directional wavelets, filamentary structures can be produced without changing the general shape of the angular power spectrum of the input FBM realization. The filamentary structures are formed through the product of a large number of random-phase linear waves at different spatial wavelengths~(see their Eq.~7). 
To extend {\tt{pywavan}} to the sphere, we made use of the {\tt{S2WAV}} Python package\footnote{\url{https://github.com/astro-informatics/s2wav}} that computes wavelet transforms on the sphere using {\tt JAX}~\citep{Price2024a, Price2024b}.\footnote{The complete Python routine used in this work, {\tt{dirade}$\_$\tt{hpx}} (DIrectional RAndom cascaDE in HEALPix), can be found at \url{http://github.com/abracco/cosmicodes/blob/master/4GMIMS/Planck_routines.py}.}

In the top boxes of \figref{fig:modFil}, we show one synthetic all-sky filamentary model obtained at $N_{\rm side} = 128$ and FWHM=80$\arcmin$ with nine distinct wavelet directions and a FBM input power spectrum of slope~-3, typical of \HI~data at intermediate and high Galactic latitudes~\citep[e.g.,][]{mamd2007, Marchal2021}. The model is shown both in a Mollweide projection~(left) and in an orthographic projection centered around the two Galactic poles~(right). In the bottom boxes, including the 20\%~mask, we show the corresponding magnetic-field orientation that was derived as follows.

We considered the relation between magnetic-field orientations perfectly aligned with density filamentary structures and the respective $TEB$~cross-correlations. In particular, perfect alignment establishes maximal $TE$~correlation without $TB$~correlation~\citep[e.g.,][]{Zaldarriaga2001, PlanckXXXVIII2016, Bracco2019a, Bracco2019b, Huffenberger2020}. Given the filamentary sky model used as a proxy of~$T$ in arbitrary units, we imposed the corresponding $E$ and $B$~modes as $E = T$ and $B = 0$, respectively. We converted this $TEB$~system to a Stokes~$IQU$ group of maps following standard relations with spherical harmonics~\citep[e.g.,][]{Bracco2019a}. We used the Python routine called {\tt{map}$\_$\tt{teb2iqu}}, which can be found at the same address written above. The modeled Stokes~$IQU$ provided us with the polarization angle and the corresponding magnetic field orientation that perfectly follows the projected morphology of the filamentary sky model. Finally, using a rotation similar to \eqref{eq:rot}, we introduced two uniform misalignment angles in the two hemispheres, namely, $\Delta \psi_{\rm N}$ and~$\Delta \psi_{\rm S}$. This last step mimicked the effect of a large-scale hemispherical misalignment between the filamentary structures, which in this model corresponds to the magnetic field traced by the \HI~templates, and the total polarization field, which corresponds to the \Planck\ polarization data. Through this synthetic filamentary model we tested the impact of~$\Delta \psi_{\rm N}$ and~$\Delta \psi_{\rm S}$ on the dust polarization power spectra as described in \secref{ssec:powerspectra_model} and shown in \figref{fig:modSpec}.

\end{document}